\documentclass{aa}
\usepackage{graphicx}
\usepackage{natbib}
\usepackage{amssymb}
\bibpunct{(}{)}{;}{a}{}{,}
\begin{document}
  \title{Does IRAS 16293--2422 have a hot core? Chemical inventory and
   abundance changes in its protostellar environment}

   \author{F. L. Sch\"{o}ier\inst{1}  \and 
           J. K. J{\o}rgensen\inst{1}   \and 
	   E. F. van Dishoeck\inst{1} \and 
	   G. A. Blake\inst{2}}

   \offprints{F. L. Sch\"{o}ier \\ \email{fredrik@strw.leidenuniv.nl}}

   \institute{Leiden Observatory, P.O. Box 9513, NL-2300 RA Leiden, 
The Netherlands   
             \and
	     Division of Geological and Planetary Sciences, 
California Institute of Technology, MS
                     150-21, Pasadena, CA 91125, USA}
	     
\date{A\&A accepted}

   \abstract{ A detailed radiative transfer analysis of the observed
     continuum and molecular line emission toward the deeply embedded
     young stellar object \object{IRAS 16293--2422} is performed.
     Accurate molecular abundances and abundance changes with radius
     are presented.  The continuum modelling is used to constrain the
     temperature and density distributions in the envelope, enabling
     quantitative estimates of various molecular abundances. The
     density structure is well described by a single power-law falling
     off as $r^{-1.7}$, i.e., in the range of values predicted by
     infall models.  A detailed analysis of the molecular line
     emission  strengthens the adopted physical model and
     lends further support that parts of the circumstellar
     surroundings of \object{IRAS 16293--2422} are in a state of
     collapse.  The molecular excitation analysis reveals that the
     emission from some molecular species is well reproduced assuming
     a constant fractional abundance throughout the envelope. The
     abundances and isotope ratios are generally close to typical
     values found in cold molecular clouds in these cases, and there
     is a high degree of deuterium fractionation.  There are, however,
     a number of notable exceptions. Lines covering a wide range of
     excitation conditions indicate for some molecules, e.g., H$_2$CO,
     CH$_3$OH, SO, SO$_2$ and OCS, a drastic increase in their
     abundances in the warm and dense inner region of the
     circumstellar envelope. The location at which this increase
     occurs is consistent with the radius at which ices are expected
     to thermally evaporate off the grains. In all, there is strong
     evidence for the presence of a `hot core' close to the protostar,
     whose physical properties are similar to those detected towards
     most high mass protostars except for a scaling factor.  However,
     the small scale of the hot gas and the infalling nature of the
     envelope lead to very different chemical time scales between low
     mass and high mass hot cores, such that only very rapidly
     produced second-generation complex molecules 
     can be formed in \object{IRAS 16293--2422}.
     Alternatively, the ices may be liberated due to grain-grain
     collisions in turbulent shear zones where the outflow interacts
     with the envelope. Higher angular resolution observations are
     needed to pinpoint the origin of the abundance enhancements and
     distinguish these two scenarios. The accurate molecular
     abundances presented for this low-mass protostar serve as a
     reference for comparison with other objects, in particular
     circumstellar disks and comets.  

      }
     \titlerunning{Does IRAS 16293--2422 have a hot core?} \maketitle
%

\section{Introduction}
Low mass protostars in their earliest stages of evolution are deeply
embedded in large amounts of dust and gas.  The nature of the emission
from such star-forming regions makes them ideal to study in the
infrared to radio wavelength regime. In the last decade, the
sensitivity of receivers operating at these wavelengths has increased
dramatically and the resulting high quality observations, supplemented
by careful modelling, have provided most of the current day knowledge
about the chemistry and physics, and their rapidly changing
properties, in early stellar evolution.  While the general scenario of
low mass stellar evolution is reasonably well understood (e.g., Shu
et~al.\ 1993\nocite{Shu93}; Evans 1999\nocite{Evans99}; Andr{\' e}
et~al.\ 2000\nocite{Andre00}) many uncertainties remain.  For example,
the nature of the hot ($T$$\gtrsim$90\,K) and dense ($n_{{\mathrm
H}_2}$$\gtrsim$10$^7$\,cm$^{-3}$) regions of gas observed towards some
low mass protostars is not yet fully established and the possible link
with the so-called hot cores observed towards most high mass
protostars needs to be investigated further.  

In the case of high mass star formation, it has become clear that hot
cores represent one of the earliest phases
\citep{Walmsley92}. Chemically, hot cores are characterized by high
abundances of fully hydrogenated molecules such as water (H$_2$O), ammonia
(NH$_3$) and hydrogen sulfide (H$_2$S), along with a rich variety of complex organic
molecules ranging from methanol (CH$_3$OH) and ethanol (CH$_3$CH$_2$OH) 
to methyl cyanide (CH$_3$CN), dimethyl ether (CH$_3$OCH$_3$), 
methyl formate (HCOOCH$_3$) and ethyl cyanide (C$_2$H$_5$CN)
 \citep{Walmsley93, Kuan96, Hatchell98b, Schilke00}.  
The chemical richness is explained
by evaporation of the ice mantles above $\sim$90~K, followed by rapid
gas-phase ion-molecule reactions leading to more complex species for a
period of $\gtrsim$10$^4$ yrs (Charnley et al.\
1995\nocite{Charnley95};  Millar et al.\ 1997\nocite{Millar97};
Rodgers \& Charnley 2001\nocite{Rodgers01},
see van Dishoeck \& Blake 1998\nocite{Dishoeck98} and Langer et al.\
2000 \nocite{Langer00} for overviews).  Low mass protostars are less
luminous and less massive, but a similar physical structure is
expected except for a scale factor \citep{Ceccarelli96, Ivezic97}.  On
the other hand, shocks due to the interaction of the outflows with the
envelope can also liberate ice mantles and drive a high-temperature
chemistry; such shocks may be relatively more important for
low-luminosity objects than for high-mass protostars (e.g., van
Dishoeck et al.\ 1995).  It is of considerable interest to establish
if low mass protostars also have hot and dense regions, and if so,
whether a similarly complex organic chemistry to that found in the
case of high-mass protostars has ensued and whether passive heating by the
accretion luminosity or active shocks dominate the liberation of grain
mantles.  Since it is the material in the warm inner envelope that
will be incorporated into circumstellar disks, it is important to know
the level of chemical complexity as it relates to forming planetary
systems.

 \object{IRAS 16293--2422} is by far the best candidate for investigating a
low mass hot core (e.g., Blake et~al.\ 1994\nocite{Blake94};
van Dishoeck et~al.\ 1995\nocite{Dishoeck95};
Ceccarelli et~al.\ 2000a,b\nocite{Ceccarelli00a}\nocite{Ceccarelli00b}).
\object{IRAS 16293--2422} is a deeply embedded low mass protostellar
object located within the $\rho$ Ophiuchus molecular cloud
complex. Due to the relative proximity of this source
\citep[160\,pc;][]{Whittet74} a wealth of molecular lines has been
detected, in spite of its relatively low luminosity
\citep[27\,L$_{\sun}$;][]{Mundy86}, and this has made \object{IRAS
16293--2422} one of the best studied young stellar objects.
Interferometer observations of radio and millimetre continuum emission
reveal two compact sources in the center of its circumstellar envelope
\citep{Wootten89, Mundy90, Mundy92, Looney00}, likely to be accretion
disks through which matter is fed onto the central stars.
 The separation of the two protostars
is approximately 800\,AU \citep{Looney00}.
\object{IRAS 16293--2422} is thought to be in one of the earliest
stages of formation; the observed spectral energy distribution (SED)
can be fitted by a modified blackbody of $\sim$40\,K
\citep[e.g.,][]{Walker86, Andre00} and has a high ratio of
submillimetre to bolometric luminosity, suggesting a large amount of
envelope mass. This places \object{IRAS 16293--2422} among a family of
deeply embedded and recently formed hydrostatic stellar objects known
as, in the traditional evolutionary sequence of low mass protostellar
objects (e.g., Andr{\' e} et al.\ 1993\nocite{Andre93}), `class 0'
protostars.

The circumstellar surroundings of this protobinary star were
extensively studied in a large molecular line survey presented in
\citet{Blake94} and \citet{Dishoeck95}. It was found that molecular
line emission is potentially a very powerful tool to probe both the
physics and chemistry of the circumstellar environment; however, a
full radiative transfer analysis was not performed and the derived
abundances have significant uncertainties. Due to the complexity of
molecular excitation and its sensitivity to the environment, various
species ---and even different lines of the same species--- probe
different parts of the circumstellar material. At least three
physically and chemically distinct parts were identified including a
circumbinary envelope, circumstellar disk(s), and outflow components.
The latter component was thought to be a small and warm region of a
few arcsec in size where the bipolar outflow(s) interact with the inner
part of the circumbinary envelope.  Recently, Ceccarelli et~al.\
(2000a,b)\nocite{Ceccarelli00a} \nocite{Ceccarelli00b} used deep JCMT
observations of H$_2$CO combined with a physical-chemical collapse
model to argue that \object{IRAS 16293--2422} does in fact have a
hot-core-like region in which the liberation of ices is consistent
with heating by the accretion luminosity.

We present here spherically-symmetric radiative transfer modelling of
the dust and gas components constituting the material in the
circumstellar envelope of \object{IRAS 16293--2422}.  The dust
parameters are constrained by the observed continuum emission in the
form of submillimetre brightness maps and the SED over a large
wavelength region. The resulting temperature and density structures
are a prerequisite to chemical studies of the molecular gas present in
the envelope. The approach taken here is different from that
adopted by Ceccarelli et~al.\ (2000a,b)\nocite{Ceccarelli00a}
\nocite{Ceccarelli00b} in that the physical parameters of the envelope
are derived empirically from the analysis of the dust emission.  Once
the physical structure of the envelope is known, a detailed excitation
analysis of molecular millimetre line emission is performed aimed at
obtaining accurate abundances. This provides valuable insight into the
complex chemistry occurring in this proto-stellar envelope. In
particular, the derived abundances allow for direct
comparison with other sources and comets (e.g., Bockel{\' e}e-Morvan
et~al.\ 2000\nocite{Bockelee00}). Moreover, searches for evidence of
abundance changes, e.g., due to evaporation of ices (`jump models') is
of considerable interest.  In addition to providing constraints on the
chemistry, the molecular line observations give further information on
the physical structure, e.g., kinematic information.  Similar
strategies have been adopted by \citet{Tak99, Tak00} and
\citet{Hogerheijde00b} and have proven to be powerful tools when
determining the physics and chemistry of star-forming regions.

\section{Observations and data reduction}
\label{observations}
In this Section the observational constraints used in the radiative
transfer modelling of the circumstellar envelope around \object{IRAS
16293--2422} are presented. The SED and submillimetre continuum
brightness maps constrain the physical structure of the
envelope. Millimetre molecular line observations provide further
information on the physical structure, in particular the large scale
velocity field, and allow for studies of the chemistry in the
envelope.

\begin{table}
\caption[]{The spectral energy distribution of \object{IRAS 16293--2422}.}
\label{SED_tab}
     $$ 
         \begin{array}{ccccccc}
            \hline
            \noalign{\smallskip}
	    \multicolumn{1}{c}{\lambda} &
	    \multicolumn{1}{c}{{F_\mathrm{obs}} ^{\mathrm a}} &
	    \multicolumn{1}{c}{{\Delta F_\mathrm{obs}} ^{\mathrm b}}  &
	    \multicolumn{1}{c}{{\theta_{\mathrm{mb}}} ^{\mathrm c}} &
	    &
	    \multicolumn{1}{c}{{F_\mathrm{mod}} ^{\mathrm d}} 
	     \\
	    
	    \multicolumn{1}{c}{[\mu\mathrm{m}]} &
	    \multicolumn{1}{c}{[\mathrm{Jy}]} &
	    \multicolumn{1}{c}{[\mathrm{Jy}]} &
	    \multicolumn{1}{c}{[\arcsec]} &
	    \multicolumn{1}{c}{\mathrm{Ref.}} &
	    \multicolumn{1}{c}{[\mathrm{Jy}]} 
	    \\
            \noalign{\smallskip}
            \hline
            \noalign{\smallskip}
	  2900& \phantom{000}0.60           & \phantom{00}0.13           & \phantom{0}60.0 & 3 & \phantom{00}0.58  \\
          1300& \phantom{000}6.97           & \phantom{00}2.24           & \phantom{0}30.0 & 3 & \phantom{00}7.5\phantom{0}  \\
\phantom{0}850& \phantom{00}30.8\phantom{0} & \phantom{00}6.2\phantom{0} & \phantom{0}15.0 & 1 & \phantom{0}31.3\phantom{0}  \\
\phantom{0}450& \phantom{0}220.4\phantom{0} & \phantom{0}44.5\phantom{0} & \phantom{00}8.8 & 1 &           220.3\phantom{0}  \\
\phantom{0}100&           1032.0\phantom{0} &           226.0\phantom{0} &           237.0 & 2 &           852.4\phantom{0}  \\
\phantom{00}60& \phantom{0}254.9\phantom{0} & \phantom{0}59.5\phantom{0} &           160.0 & 2 &           162.4\phantom{0}  \\
            \noalign{\smallskip}
            \hline
         \end{array}
     $$
     \smallskip

     \noindent
     $^{\mathrm a}$ Observed flux integrated over emitting region. \\
     \noindent
     $^{\mathrm b}$ A calibration uncertainty of 20\% has been added \\
     \noindent
     $^{\mathrm c}$ Size of the main beam. \\
     \noindent
     $^{\mathrm d}$ Flux predicted by best fit model using a density structure described by a single power-law and
     OH5 dust opacities. Note that the 60\,$\mu$m and 2.9\,mm fluxes are not included in the modelling (see text for details). \\
     \noindent
     Refs. -- (1) This paper;  (2) IRAS Point Source Catalogue; (3) \citet{Walker90}.

   \end{table}
   \begin{figure} 
   \includegraphics[width=88mm]{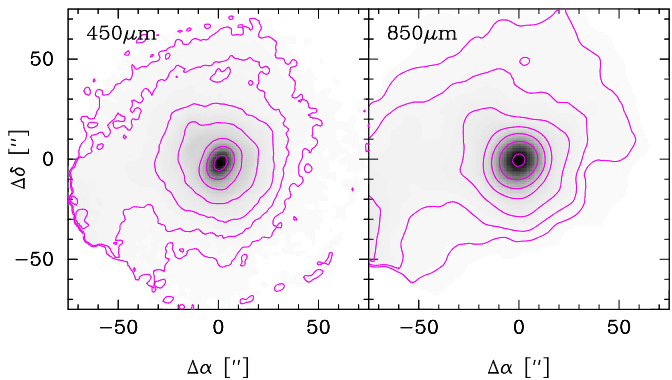}
   \caption{SCUBA images at 450 and 850\,$\mu$m of \object{IRAS
   16293--2422}. The contours start at the 3$\sigma$ level (0.9 and
   0.24 Jy\,beam$^{-1}$ for 450 and 850\,$\mu$m, respectively) and
   increase by multiples of 2.}  
   \label{maps}
   \end{figure}

\subsection{Spectral energy distribution}
The SED of \object{IRAS 16293--2422}, as presented in
Table~\ref{SED_tab} is used to constrain the total amount of material
present in the circumstellar envelope.  Care has been taken that only
total fluxes integrated over the whole emitting region for wavelengths
$\leq$1.3 millimetre (mm) are used in the SED modelling to ensure that emission
from the envelope itself is dominating the observed flux. The
contribution from the disk(s) to the observed emission starts to become
important at wavelengths longer than $\sim$1\,mm. However, any disk emission 
would be significantly diluted in the relatively large beams used here 
(e.g., Motte \& Andr{\' e} 2001\nocite{Motte01}).
That thermal emission from an extended dusty envelope is responsible for the
majority of the observed millimetre flux is confirmed by the radiative transfer
analysis performed in Sect.~\ref{dustres}. The model fluxes predicted by the best 
fit envelope model are compared with observations in Table~\ref{SED_tab}.
Additionally, the observed IRAS flux at 60\,$\mu$m is not used in the
analysis since the emission at that wavelength emanates from the inner
hot parts of the envelopes where the dust grain properties are 
probably significantly different from those in the cooler outer parts.
Dust grains in the outer parts will be coated by a layer of ice which,
as the temperature increases towards the star, starts to evaporate
thereby changing the optical properties of the dust grains.  In
Sect.~\ref{dustres}, the effect of varying the dust opacities will be
investigated.
A relative calibration uncertainty of 20\% is added
to all flux measurements, in addition to any statistical errors, which
dominates the error budget and gives all 
 points on the SED more or less equal weights.

\subsection{Submillimetre continuum observations}
Submillimetre continuum observations were retrieved from the James
Clerk Maxwell Telescope (JCMT) public
archive\footnote{{\tt http://www.jach.hawaii.edu/JACpublic/JCMT/} 
\\
The JCMT is operated by the Joint Astronomy Centre in Hilo, Hawaii on behalf of the
present organizations: the Particle Physics and Astronomy Research Council in the 
United Kingdom, the National Research Council of Canada and the Netherlands
Organization for Scientific Research.}.  The
data were obtained during an observational run in April 1998, using
the Submillimetre Common-User Bolometer Array (SCUBA), and consist of
pairs of images at 450\,$\mu$m and 850\,$\mu$m. The dual SCUBA array
contains 91 pixels in the short-wavelength array and 37 pixels in the
long-wavelength array, each covering a hexagonal 2$\farcm$3 field.  The
SCUBA bolometer array camera is described in some detail in
\citet{Holland99}.

The imaging was made using the jiggle-mapping mode to produce fully
sampled maps. In this mode the SCUBA bolometers instantaneously
under-sample the sky and a 64 point jiggle pattern is carried out by
the telescope to fully sample both the long and short wavelength
arrays.  In practice, the 64 point pattern is broken down into four
16-point sub-patterns that spend 1\,s integrating on the source at
each gridpoint. After a sub-pattern has been completed the telescope
is nodded and the pattern is repeated again so that sky subtraction
can be made. In all, it takes 128\,s to complete one full on/off
source jiggle map. The jiggle mapping mode is the preferred
observational mode for SCUBA when imaging sources smaller than the
chop throw. Usually, a chop throw of 2$\arcmin$ is used to ensure
chopping off the arrays. A larger chop throw would result in poor sky
subtraction and loss in image quality.

\begin{table}
\caption[]{Observational results from SCUBA images.}
\label{obs}
     $$ 
\begin{array}{cccccc}
\hline
\noalign{\smallskip}
\multicolumn{1}{c}{\lambda} &
\multicolumn{1}{c}{{\tau} ^{\mathrm a}} &
\multicolumn{1}{c}{{F_\mathrm{tot}} ^{\mathrm b}} &
\multicolumn{1}{c}{{F_\mathrm{peak}} ^{\mathrm c}} &
\multicolumn{1}{c}{\mathrm{rms}} &
\multicolumn{1}{c}{{\theta_{\mathrm{mb}}} ^{\mathrm d}}\\

\multicolumn{1}{c}{[\mu\mathrm{m}]} &
&
\multicolumn{1}{c}{[\mathrm{Jy}]} &
\multicolumn{1}{c}{[\mathrm{Jy\,beam}^{-1}]} &
\multicolumn{1}{c}{[\mathrm{Jy\,beam}^{-1}]} & 
\multicolumn{1}{c}{[\arcsec]}\\
\noalign{\smallskip}
\hline
\noalign{\smallskip}
450 & 0.5\phantom{0} - 0.7\phantom{0} & 220.4		 & 72.0 & 0.3\phantom{0} & \phantom{0}8.8\\
850 & 0.12 - 0.15		      & \phantom{0}30.8  & 16.8 & 0.08 & 15.0 \\
\noalign{\smallskip}
\hline
\end{array}
$$ 
\smallskip

\noindent
$^{\mathrm a}$ Zenith opacity of the atmosphere.\\
$^{\mathrm b}$ Total flux integrated over emitting region.
	       The calibration uncertainty is estimated to be $\sim$20\%.\\
$^{\mathrm c}$ Flux at stellar position.
	       The calibration uncertainty is estimated to be $\sim$20\%.\\
$^{\mathrm d}$ Geometrical mean of the beam size.\\
\end{table}

\begin{table*}
\caption[]{Additional molecular line observations of \object{IRAS 16293--2422} using the JCMT.}
\label{radio}
$$ 
\begin{array}{p{0.1\linewidth}cccccc}
\hline
\noalign{\smallskip}
&
&
\multicolumn{1}{c}{\mathrm{Frequency}} &
\multicolumn{1}{c}{{\int T_{\mathrm{mb}} dV} ^{\mathrm a}} &
\multicolumn{1}{c}{{T_{\mathrm{mb}}} ^{\mathrm b}} &
\multicolumn{1}{c}{{\Delta V} ^{\mathrm b}} &
  \\

\multicolumn{1}{c}{\mathrm{Molecule}} &
\multicolumn{1}{c}{\mathrm{Line}} &
\multicolumn{1}{c}{[\mathrm{MHz}]} &
\multicolumn{1}{c}{[\mathrm{K\,km\,s}^{-1}]} &
\multicolumn{1}{c}{[\mathrm{K}]} &
\multicolumn{1}{c}{[\mathrm{km\,s}^{-1}]} &
\multicolumn{1}{c}{\mathrm{Ref.}} \\
\noalign{\smallskip}
\hline
\noalign{\smallskip}
$^{13}${CO}		  & J=2-1\phantom{00}	&220401.7 &	      50.0\phantom{0} & \phantom{0}9.7\mathrm{:}\phantom{0}  & 3.5\mathrm{:}  & 1\\
			  & J=3-2\phantom{00}	&330588.1 &	      67.2\phantom{0} & \phantom{0}8.5\mathrm{:}\phantom{0}  & 4.7\mathrm{:}  & 1\\
C$^{18}$O		  & J=2-1\phantom{00}	&219560.4 &	      21.6\phantom{0} & \phantom{0}7.8\phantom{:}\phantom{0} & 2.6\phantom{:} & 1\\
			  & J=3-2\phantom{00}	&329335.0 &	      33.6\phantom{0} & 	  10.1\phantom{:}\phantom{0} & 3.2\phantom{:} & 1\\
C$^{17}$O		  & J=2-1\phantom{00}	&224714.4 & \phantom{0}6.6\phantom{0} & \phantom{0}2.5\phantom{:}\phantom{0} & 2.7\phantom{:} & 1\\
			  & J=3-2\phantom{00}	&337061.1 &	      10.9\phantom{0} & \phantom{0}3.0\phantom{:}\phantom{0} & 3.4\phantom{:} & 1\\
CS			  & J=5-4\phantom{00}	&244935.7 &	      41.1\phantom{0} & 	  11.1\phantom{:}\phantom{0} & 3.3\phantom{:} & 1\\
			  & J=7-6\phantom{00}	&342887.8 &	      51.4\phantom{0} & 	  13.6\phantom{:}\phantom{0} & 3.3\phantom{:} & 1\\
C$^{34}$S		  & J=5-4\phantom{00}	&241016.2 & \phantom{0}4.8\phantom{0} & \phantom{0}1.3\phantom{:}\phantom{0} & 3.4\phantom{:} & 1\\
			  & J=7-6\phantom{00}	&342883.0 & \phantom{0}5.7\phantom{0} & \phantom{0}1.6\phantom{:}\phantom{0} & 3.2\phantom{:} & 1\\
HCN			  & J=3-2\phantom{00}	&265886.4 &	      51.3\phantom{0} & \phantom{0}4.0\mathrm{:}\phantom{0}  & 7.5\mathrm{:}  & 1\\
			  & J=4-3\phantom{00}	&354505.5 &	      63.4\phantom{0} & \phantom{0}5.0\mathrm{:}\phantom{0}  & 7.3\mathrm{:}  & 1\\
HNC			  & J=3-2\phantom{00}	&271981.1 &	      14.2\phantom{0} & \phantom{0}4.7\mathrm{:}\phantom{0}  & 2.9\mathrm{:}  & 1\\
			  & J=4-3\phantom{00}	&362630.1 &	      11.9\phantom{0} & \phantom{0}4.1\mathrm{:}\phantom{0}  & 3.0\mathrm{:}  & 1\\
HCO$^{+}$		  & J=3-2\phantom{00}	&267557.6 &	      70.3\phantom{0} & 	  18.0\mathrm{:}\phantom{0}  & 3.5\mathrm{:}  & 1\\
			  & J=4-3\phantom{00}	&356734.0 &	      95.9\phantom{0} & 	  21.5\mathrm{:}\phantom{0}  & 4.0\mathrm{:}  & 1\\
H$^{13}$CO$^{+}$	  & J=3-2\phantom{00}	&260255.5 &	      10.3\phantom{0} & \phantom{0}3.8\mathrm{:}\phantom{0}  & 2.6\mathrm{:}  & 2\\
			  & J=4-3\phantom{00}	&346998.5 & \phantom{0}8.5\phantom{0} & \phantom{0}3.1\phantom{:}\phantom{0} & 2.7\phantom{:} & 1\\
$^{29}$SiO		  & J=8-7\phantom{00}	&342979.1 & \phantom{0}1.8\phantom{0} & \phantom{0}0.3\phantom{:}\phantom{0} & 5.4\phantom{:} & 1\\
HC$_3$N 		  & J=24-23 &218324.8 & \phantom{0}0.64 	  & \phantom{0}0.15\phantom{:}  & 3.9\phantom{:}  & 1\\
\noalign{\smallskip}
\hline
\end{array}
$$ 
\smallskip

\noindent
$^{\mathrm a}$ Total integrated intensity calculated over full extent of line. 
	       The calibration uncertainty in the intensity scale is estimated to be $\sim$15$-$20\%.\\
\noindent
$^{\mathrm b}$ Estimated from a gaussian fit to the observed spectrum.  A colon (:)
   	    indicates an uncertain value due to low signal-to-noise or, in the majority of cases, 
   	    a significant departure from a gaussian line profile.\\
\noindent
Refs. -- (1) JCMT public archive; (2) This paper.
\end{table*}

The data were reduced in a standard way, as described in
\citet{Sandell97}, using the SCUBA reduction package SURF
\citep{Jenness97}. The images were calibrated using simultaneous
observations of Uranus retrieved from the JCMT archive.  The sky
opacities at 450\,$\mu$m and 850\,$\mu$m were estimated using the
1.3\,mm opacity, monitored by the Caltech Submillimeter Observatory
and listed for each individual observation, using the relations in
\citet{Archibald00}.  The validity of these relations have been
checked and confirmed using SCUBA sky dips. The total calibration
uncertainty is estimated to be approximately $\pm$20\% 
at 450\,$\mu$m and about  $\pm$10\% at 850\,$\mu$m.  
Care was
taken to select data taken during good to excellent submillimetre
conditions.  The beam is determined from the Uranus observations and
is dominated by an approximately gaussian main beam with a deconvolved
FWHM of 14$\farcs$2$\times$16$\farcs$0 and
8$\farcs$5$\times$9$\farcs$1 at 850\,$\mu$m and 450\,$\mu$m,
respectively. A substantial error beam is, however, present at these
wavelengths (see below) picking up significant amounts of flux. The
error lobe pick up is estimated to be approximately 15\% and 45\% at
850\,$\mu$m and 450\,$\mu$m, respectively, and is taken explicitly
into account in the analysis.

The final 450\,$\mu$m and 850\,$\mu$m images are presented in
Fig.~\ref{maps}.  All offsets reported are relative to the adopted
position of the protobinary star \object{IRAS 16293--2422}
($\alpha_{2000}$$=$16$^{\mathrm h}$32$^{\mathrm m}$22\fs 91,
$\delta_{2000}$$=$$-$24\degr 28\arcmin 35\farcs 6).  The locations of
the two protostars IRAS 16293A (MM1) and IRAS 16293B (MM2) relative to
this position are ($-$3$\arcsec$, $-$1$\arcsec$) and ($-$5$\arcsec$,
$+$3$\arcsec$), respectively.  The emission appears to have an overall
spherical symmetry and is centered on the adopted central position
within the pointing accuracy of the telescope.  For comparison, the
pointing accuracy of the JCMT is estimated to be about
$\pm$1.5$\arcsec$ in both elevation and azimuth.  Also visible is a
second, weak, component apparent near the eastern edge of the SCUBA
maps.  This second component, or IRAS 16293E ($+$77$\arcsec$, $-$22$\arcsec$),
which was first identified by its strong ammonia emission, is most
probably also a class 0 protostar \citep{Mizuno90, Castets01}.

The FWHM of the emission centered on the stellar position is
20$\farcs$8$\times$19$\farcs$4 and 21$\farcs$9$\times$19$\farcs$3 at
850\,$\mu$m and 450\,$\mu$m, respectively. The deconvolved envelope
sizes assuming both the beam and brightness distribution to be
described by gaussian functions are $\sim$14$\arcsec$ at 850\,$\mu$m
and $\sim$19$\arcsec$ at 450\,$\mu$m.  Thus, only the 450\,$\mu$m
emission appears to be resolved.  The observational results are
summarized in Table~\ref{obs}.  To compare the observed brightness
distributions with the predictions from a spherically symmetric model,
the SCUBA maps were azimuthally averaged in bins with half the
corresponding beam size in width.  Moreover, care was taken to block
out any contribution from IRAS 16293E.  The resulting radial brightness
distributions are shown in Fig.~\ref{bestfit_dust}.

\subsection{Millimetre molecular line observations}
A survey of the millimetre molecular line emission towards
\object{IRAS 16293--2422} was presented in \citet{Blake94} and
\citet{Dishoeck95}. This large data set forms the base for the
molecular excitation analysis performed in this paper. The absolute
calibration uncertainty of the intensities is estimated to be
$\sim$30\%.  In addition, we have searched the JCMT public archive for
complementary millimetre line observations.  This additional set of
data, taken at face value, is presented in Table~\ref{radio}.
Lines for which multi-epoch observations are available in the
JCMT archive typically display intensities that are
consistent to $\sim$20\%.  This was also the conclusion reached by
\citet{Schoeier00} for a large survey of carbon stars.  When newer
data were available for a particular transition, they were usually
adopted, due to a higher signal-to-noise and/or greater spectral
resolution. In one case [H$^{13}$CO$^{+}$($J$$=$3$\rightarrow$2)]
the old data set was found to have a significantly lower line
intensity, possibly due to pointing problems.
Additional H$_2$CO data published recently by 
\citet{Loinard00} were further used in order to increase the number of 
observed transitions for this molecule. 

The detected molecular line emission probes the full radial range of
the envelope, providing additional constraints on the physical
structure of \object{IRAS 16293--2422}.  Only information on the lowest
transitions of the molecules, which occur at millimetre wavelengths
and probe the very coldest outer parts, is lacking. The observed line
shapes provide valuable information on the velocity structure in the
envelope.
For example, the single-dish observations of abundant molecules 
like CO and CS typically show lines with
strong self-absorption and some degree of asymmetry.  The variation
of the line profiles among the CO and CS isotopomers is potentially
a sensitive probe of infall models and will be further
investigated in Sect.~\ref{infall}.

\section{Radiative transfer}
In this Section the continuum and molecular line radiative transfer
codes used to constrain the physical and chemical structure of the
envelope around \object{IRAS 16293--2422} are presented. The envelope
is assumed to be spherically symmetric and the approach adopted here
is to first determine the density and temperature structures from the
dust modelling.  
This, in turn, allows for abundances of various
molecules present in the circumstellar envelope to be determined.

\subsection{Dust radiative transfer model}
\label{dustmod}
In order to model the observed continuum emission and to be able to
extract some basic parameters of the dusty envelope around
\object{IRAS 16293--2422} the publically available dust radiative
transfer code DUSTY\footnote{\tt{http://www.pa.uky.edu/\~{
}moshe/dusty/}} \citep{Ivezic99} has been adopted.  DUSTY makes use of
the fact that in some very general circumstances the radiative
transfer problem of the dust possesses scaling properties
\citep{Ivezic97}. The solution is presented in terms of the distance,
$r$/$r_{\mathrm{i}}$, scaled with respect to the inner boundary
$r_{\mathrm{i}}$.  In addition to the properties of the dust and the
relative size of the envelope, the only parameter needed for a full
description of the problem is the spectral shape of the radiation
emitted by the central source.  This means that the luminosity is
totally decoupled from the radiative transfer problem and it is only
used to scale the solution in order to obtain the absolute distance
scale.  This scaling property of the dust radiative transfer is very
practical, in particular when modelling a large number of similar
sources, and is put to use in \citet{Jorgensen02} for a survey of
low mass protostars at various evolutionary stages  and in 
\citet{Hatchell00} to model a selection of high mass protostars.

The most important parameter controlling the output is the dust optical depth
\begin{equation}
\label{tau1}
\tau_{\lambda} = \kappa_{\lambda} \int_{r_{\mathrm i}}^{r_{\mathrm e}} \rho_{\mathrm d}(r)\,dr,
\end{equation}
where $\kappa_{\lambda}$ is the dust opacity, $\rho_{\mathrm d}$ the
density distribution of the dust in the envelope, and $r_{\mathrm i}$
and $r_{\mathrm e}$ the inner and outer radius of the dust envelope,
respectively. Introducing the dust-to-gas mass ratio, $\delta$, allows
Eq.~\ref{tau1} to be written
\begin{equation}
\label{tau2}
\tau_{\lambda} = \kappa_{\lambda} \delta \mathrm{m_{H_2}} 
                 \int_{r_{\mathrm i}}^{r_{\mathrm e}} n_{\mathrm H_2}(r)\,dr = 
                 \kappa_{\lambda} \delta \mathrm{m_{H_2}} N_{\mathrm H_2},
\end{equation}
where $n_{\mathrm H_2}$ is the number density distribution of
molecular hydrogen, $\mathrm{m_{H_2}}$ the mass of an hydrogen
molecule, and $N_{\mathrm H_2}$ the column density of molecular
hydrogen. In the derivation of Eq.~\ref{tau2} any possible drift
velocity between the dust and the gas has been neglected. In what
follows $\delta$=0.01 is assumed.  The dust
opacities from \citet{Ossenkopf94} were used, corresponding to
coagulated dust grains with thin ice mantles, at a density of
$n_{\mathrm H_2}$$\sim$10$^6$\,cm$^{-3}$ (column 5 in their Table~1,
hereafter OH5).  Van der Tak et~al.\ (1999)\nocite{Tak99} considered
various sets of dust optical properties when modelling the high mass
young stellar object \object{GL 2591}, and found that models using the
OH5 opacities were the only ones that gave envelope
masses consistent with those derived from the modelling of the
molecular line emission.  In Sect.~\ref{dust_oh5} another set of
opacities will also be considered, more appropriate for regions where the
ices have evaporated from the dust grains.

The dust temperature at the inner radius of the envelope is fixed to
300\,K and this sets the inner radius. The choice of this temperature
is motivated by the observations of line emission arising from highly
excited molecules in the envelope.  However, the inner regions are
complex with breakdown of spherical symmetry and interactions between
disk, envelope and outflow. In the present analysis these
complications are ignored and for simplicity a smooth and
spherically-symmetric envelope is assumed.  Recently,
\citet{Ceccarelli00a} successfully modelled molecular line emission in
the envelope around \object{IRAS 16293--2422} down to $\sim$30\,AU,
assuming spherical symmetry.  
While the separation of the two protostars
is approximately 800\,AU \citep{Looney00}, one of the protostars
IRAS 16293A (MM1)  exhibits jet-like centimetre
wavelength emission,  water maser emission and associated millimetre 
molecular emission, thus appearing to be significantly more active 
(Wootten \& Loren 1987\nocite{Wootten87}; Mundy et~al.\ 1992\nocite{Mundy92}; 
Sch\"{o}ier et~al.\ 2002a, in prep.), which justifies the approach
taken here.  The central source of radiation is assumed to arise from
a blackbody at 5000\,K. This is an oversimplification considering the
binary nature of \object{IRAS 16293--2422} and the uncertainty in the
intrinsic SED of a protostar.  The final model does not depend on 
the exact stellar temperature adopted, however, within a reasonable
range of values, since the radiation is totally reprocessed by the
circumstellar dust. The input parameters are summarized in
Table~\ref{dusty}.

The observational constraints, as presented in
Sect.~\ref{observations}, are the SED and radial brightness
distributions at 450\,$\mu$m and, to a lesser extent, 850\,$\mu$m.
The ability of the model to reproduce the observational constraints
are quantified using the chi-squared statistic
\begin{equation}
\label{chi2_sum}
\chi^2 = \sum^N_{i=1} \left [ \frac{(F_{\mathrm{mod},i}-F_{\mathrm{obs},i})}{\sigma_i}\right ]^2, 
\end{equation} 
where $F$ is the flux and $\sigma_i$ the uncertainty in observation
$i$, and the summation is done over all $N$ independent observations.
The radial brightness distributions from DUSTY are extended into 2D
surface brightness maps and convolved with the beam as determined from
planet observations. The beam convolved maps are then azimuthally
averaged in 3$\arcsec$ bins.  In the $\chi^2$ fitting procedure only
data points separated by one full beam are used since the
$\chi^2$-analysis, in practice, requires uncorrelated measurements.
In the analysis only the inner 50$\arcsec$ of the brightness distributions
are considered which are well above the background emission and should not
be significantly affected by the 120$\arcsec$ chop throw.
Furthermore, the IRAS fluxes from the model were convolved with the
proper filters before the $\chi^2$ analysis of the SED was made.  The
results from the dust radiative transfer are presented in
Sect.~\ref{dustres}.

\subsection{Molecular line radiative transfer model}
\label{gasmod}
In order to derive accurate molecular abundances for the wealth of
molecular line emission detected toward this source the detailed
non-LTE radiative transfer code of \citet{thesis}, based on the Monte
Carlo method, was used. The code produces output that is in excellent
agreement with the Monte Carlo code presented by
\citet{Hogerheijde00}. It has also been tested against other molecular
line radiative transfer codes, for a number of benchmark problems, to
a high accuracy \citep{Zadelhoff02}.

Adopting the parameters of the circumstellar envelope derived from the
dust radiative transfer analysis, the Monte Carlo code calculates the
steady-state level populations of the molecule under study, using the
statistical equilibrium equations.  In the Monte Carlo method,
information on the radiation field is obtained by simulating the line
photons using a number of model photons, each representing a large
number of real photons from all transitions considered.  These model
photons, emitted locally in the gas as well as injected from the
boundaries of the envelope, are followed through the envelope and the
number of absorptions are calculated and stored. Photons are
spontaneously emitted in the gas with complete angular and frequency
redistribution, i.e., the local emission is assumed to be isotropic
and the scatterings are assumed to be incoherent.  The weight of a
model photon is continuously modified as it travels through the
envelope, to take the absorptions and stimulated emissions into
account. When all model photons are absorbed in, or have escaped from,
the envelope the statistical equilibrium equations are solved and the
whole process is then repeated until some criterion for convergence is
fulfilled.  Once the molecular excitation, i.e., the level
populations, is obtained the radiative transfer equation can be solved
exactly.  The resulting brightness distribution is then convolved with
the appropriate beam to allow a direct comparison with observations.
In this analysis, the kinetic temperature of the gas is assumed
to follow that of the dust \citep{Ceccarelli96, Doty97, Ceccarelli00a}.

Typically, energy levels up to $\sim$500\,K in the
ground vibrational state were retained in the analysis. Vibrationally
excited levels are not included since the radiative excitation
due to the dust is generally inefficient and, 
for the temperature and 
density ranges present here, collisional excitation to these levels is
negligible.  Line emission from rotational transitions within
vibrationally excited states has, however, been observed for some
species [e.g., the CS($v$$=$1, $J$$=$7$\rightarrow$6)
\citep{Blake94}].  In Sect.~\ref{molabund} the nature of such
emission is discussed further.   Collisional rate coefficients are taken from the
literature and, in the case of some linear molecules, extrapolated
both in temperature and to transitions involving energy levels with
higher $J$ quantum numbers when needed [see Sch\"{o}ier et~al.\ (2002b, in prep.) for
details].  For other (non-linear) species for which such extrapolations
are not obvious, the excitation is assumed to be in LTE for levels for
which no collisional rate coefficients are available.

In what follows, all quoted abundances, $f_{\mathrm X}$, for a
particular molecular species X are relative to that of molecular
hydrogen, i.e.,
\begin{equation}
f_{\mathrm X}(r) = \frac{n_{\mathrm X}(r)}{n_{\mathrm{H_2}}(r)},
\end{equation}
and are initially assumed to be constant throughout the envelope.
Subsequently, it will be shown that in order to model the observed
line emission for some molecules, e.g, H$_2$CO and SiO, this latter
constraint has to be relaxed and a jump in $f$ needs to be introduced
(see also Ceccarelli et al.\ 2000a,b\nocite{Ceccarelli00a}\nocite{Ceccarelli00b}).
 Van der Tak et al.\ (2000b)\nocite{Tak00b}
 used a similar approach for the case of massive protostars.

The beam profile used in the convolution of the modelled emission is
assumed to be gaussian which is appropriate at the frequencies used
here.  The best fit model is estimated from the $\chi^2$-statistic
defined in Eq.~\ref{chi2_sum} using the observed integrated
intensities and assuming a 30\% calibration uncertainty.  Even though
the new data presented in Table~\ref{radio} appear to be slightly
better calibrated, the old data set constitutes the vast majority of
observational constraints so the larger calibration uncertainty is
applied to the full data set, for simplicity. Only in the cases of 
CO and CS was a lower calibration uncertainty
of 15\% adopted. These molecules are regularly observed towards
\object{IRAS 16293--2422} and used as standard spectra.

\section{The physical structure of the envelope}
\label{dustres}
The envelope parameters, such as the density and temperature
structures, are constrained in this Section using mainly the observed
continuum emission.

\subsection{Simple analysis}
For optically thin dust emission one can derive the following
expression for its intensity, in the Rayleigh-Jeans limit, as a
function of the impact parameter $b$ \citep{Shirley00}
\begin{equation}
\frac{I_{\nu}(b)}{I_{\nu}(0)} = \left (\frac{b}{b_0} \right )^{-\gamma} , \gamma=\alpha+\beta-1,
\end{equation}
assuming that the density 
\begin{equation}
n_{\mathrm d}(r) \propto r^{-\alpha}
\end{equation}
and temperature
\begin{equation}
T_{\mathrm d}(r) \propto r^{-\beta}
\end{equation}
follow power-law distributions.  For \object{IRAS 16293--2422}
$\gamma$$\sim$1.8 is derived in the range
$b$$\sim$15$\arcsec$$-$50$\arcsec$\ for the 450\,$\mu$m data.
Assuming $\beta$=0.4 (c.f., Doty \& Leung 1994\nocite{Doty94}; 
Shirley et al. 2000\nocite{Shirley00}; see also Fig.~\ref{tdust}), 
$\alpha$$\sim$2.4 is obtained, consistent with
the range of values determined by Shirley et al.\ for a sample of
protostellar objects.  As discussed by Shirley et al.\ there are a
number of caveats when using this simplistic approach, e.g., the
validity of applying the Rayleigh-Jeans approximation in the cool
outer parts of the envelope and deviations of the dust temperature
from a single power-law (see also Hogerheijde \& Sandell\
2000; Doty \& Palotti 2002\nocite{Doty02}). 
Shirley et~al.\ estimate the uncertainties in the derived
$\alpha$ to be of the order $\pm$0.5 using this simple approach.

\subsection{Power-law density model}
\label{dust_oh5}
In order to derive more reliable envelope parameters the dust
radiative transfer model presented in Sect.~\ref{dustmod} is used.
The density of the dust (and gas) is assumed to follow a simple
power-law
\begin{equation}
n_{\mathrm H_2}(r) = n_0 \left ( \frac{1000\,\mathrm{AU}}{r} \right ) ^{\alpha},
\end{equation}
where $n_0$ is the number density of H$_2$ at a distance of 1000\,AU
from the star.  The models where the density structure is given by a
single power-law will be referred to as static envelopes, since in the
molecular excitation analysis no large-scale velocity field is
included.  The total amount of dust and its spatial distribution are
simultaneously determined, i.e., $\tau_{100}$ (the optical depth at
100\,$\mu$m), $\alpha$, and $r_{\mathrm e}/r_{\mathrm i}$ (the
geometrical envelope thickness) are the adjustable parameters in our
model.
 Scaling the model output to the source luminosity of
27\,L$_{\sun}$ and a distance of 160\,pc fixes the absolute scale and
allows various physical parameters of the
envelope around \object{IRAS 16293--2422} to be determined.

\begin{figure}
 \includegraphics[width=88mm]{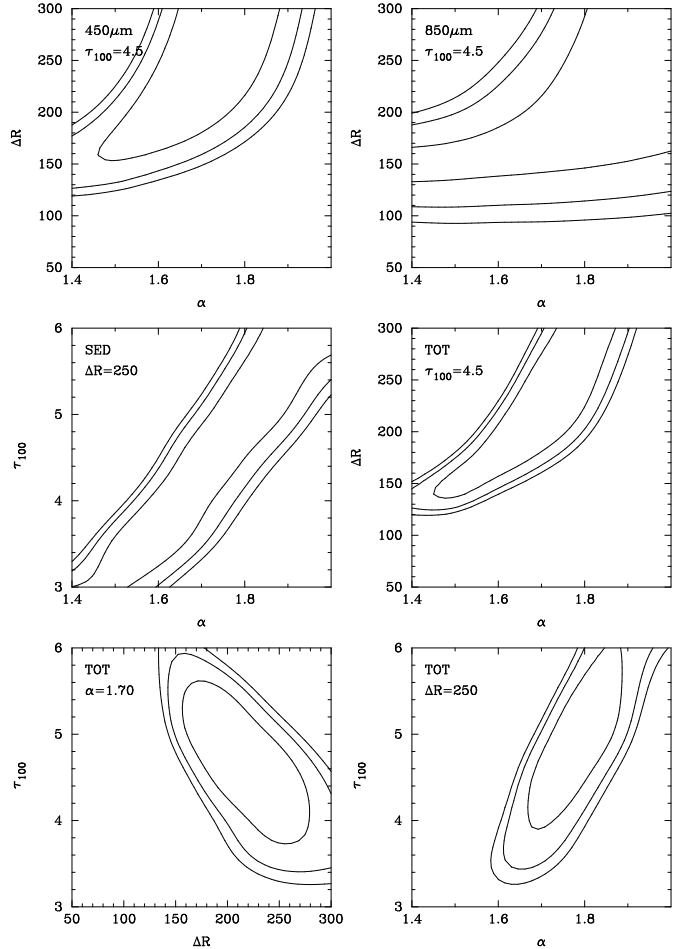}
 \caption{$\chi^2$ maps showing the sensitivity of the single 
 power-law model to the adjustable parameters 
 $\tau_{100}$, $\alpha$,
 and $\Delta R$$=$$r_{\mathrm e}$/$r_{\mathrm i}$ using the SED and the SCUBA maps as observational
 constraints. Contours 
 are drawn at $\chi^2_{\mathrm{min}}$$+$(2.3, 4.6, 6.2) 
 indicating the 68\% (`1$\sigma$'), 90\%, and 95\% (`2$\sigma$') 
 confidence levels, respectively.}
 \label{chi2maps}
\end{figure}

To find the best fit model in the 3D parameter space the strategy
adopted is to first analyze the model grid by applying the observed
radial brightness distributions. As shown in Fig.~\ref{chi2maps}, the
450\,$\mu$m and 850\,$\mu$m brightness distributions are sensitive to
the slope of the density distribution and, to a lesser extent, the
size of the dusty envelope. Changing the total amount of dust by
changing $\tau_{100}$ has only a minor effect on the allowed values of
$\alpha$ since normalized radial brightness distributions are used.
Values of $\alpha$ in the range 1.5$-$1.9 are found to be acceptable,
with a preferred value of 1.7. A typical accuracy of $\pm$0.2 in the
derived value of $\alpha$ was also found by \citet{Jorgensen02} when
analyzing SCUBA images at 450\,$\mu$m and 850\,$\mu$m for a large
sample of protostars.  In general, the 450\,$\mu$m data should provide
a more reliable value of $\alpha$ because of the higher resolution, which
provides a larger sensitivity to changes in the density structure.

The SED provides a good constraint on $\tau$ once the density profile
is known (Fig.~\ref{chi2maps}; see also see also Doty \& Palotti 2002\nocite{Doty02}). 
For a density slope of 1.7 the optical
depth at 100\,$\mu$m is estimated to be approximately 4.5.  At
450\,$\mu$m and 850\,$\mu$m the optical depths are $\sim$0.4 and
$\sim$0.1 respectively.  
For $\alpha$$\geq$1.7 the size of the envelope is not well constrained
which is not surprising given that only points in the brightness
distributions out to 50$\arcsec$ ($r_{\mathrm e}/r_{\mathrm i}$$\sim$250) are used.
The circumstellar envelope will eventually merge with the more extended cloud material in which
the object is embedded. The maximum outer radius of the envelope is fixed at the point where 
the dust temperature reaches 10\,K.
The envelope size $r_{\mathrm e}/r_{\mathrm i}$ is estimated to be 
250 for the adopted $\alpha$ of 1.7.

\begin{table}
 \caption[]{Summary of the dust radiative transfer analysis of
 \object{IRAS~16293--2422} using a single power-law density distribution (see text for details).}
 \label{dusty}
 $$
 \begin{array}{p{0.6\linewidth}r}
 \multicolumn{2}{c}{\mathrm{Fixed\ input\ parameters}} \\
 \hline
 \noalign{\smallskip}
 Distance, $d$	  & 160\,\mathrm{pc}	  \\
 Luminosity, $L$   & 27\,\mathrm{L}_{\sun}\\
 Stellar temperature, $T_{\star}$    & 5000\,\mathrm{K}\\
 Dust temperature at $r_{\mathrm{i}}$, $T_{\mathrm{d}}(r_{\mathrm{i}})$   & 300\,\mathrm{K}\\
 Dust opacity (OH5) at 100\,$\mu$m, $\kappa_{100}$ & 86.5\,\mathrm{cm}^2\mathrm{g}^{-1} \\
 \hline
 \noalign{\bigskip}
 \multicolumn{2}{c}{\mathrm{Variable\ input\ parameters}} \\
 \hline
 \noalign{\smallskip}
 Dust optical depth at 100\,$\mu$m, $\tau_{100}$     & 3.0 - 6.0\\
 Density power law index, $\alpha$   & 1.4 - 2.0 \\
 Envelope thickness, $r_{\mathrm{e}}$/$r_{\mathrm{i}}$  & 50 - 300 \\

 \hline
 \noalign{\bigskip}
 \multicolumn{2}{c}{\mathrm{Best\ fit\ parameters}} \\
 \hline
 \noalign{\smallskip}
 Dust optical depth at 100\,$\mu$m, $\tau_{100}$     & 4.5 \\
 Density power law index, $\alpha$   & 1.7 \\
 Envelope thickness, $r_{\mathrm{e}}$/$r_{\mathrm{i}}$  & 250 \\
 \hline
 \noalign{\bigskip}
 \multicolumn{2}{c}{\mathrm{Derived\ parameters}} \\
 \hline
 \noalign{\smallskip}
 Inner envelope radius, $r_{\mathrm{i}}$ 	 & 4.8\times 10^{14}\,\mathrm{cm}\\
 Outer envelope radius, $r_{\mathrm{e}}$ 	 & 1.2\times 10^{17}\,\mathrm{cm}\\
 H$_2$ column density, $N(\mathrm{H}_2)$ 	 & 1.6\times 10^{24}\,\mathrm{cm}^{-2} \\
 H$_2$ density at 1000\,AU, $n_0$		 & 6.7\times 10^{6}\,\mathrm{cm}^{-3} \\
 Envelope mass, $M_{\mathrm{env}}$		 & 5.4\,\mathrm{M}_{\sun} \\
 Bolometric flux, $F_{\mathrm{bol}}$		 & 3.4\times 10^{-11}\,\mathrm{W\,m}^{-2}  \\
 \hline
 \noalign{\smallskip}
 \end{array}
 $$	 
\end{table}
\begin{figure*} 
 \includegraphics[width=180mm]{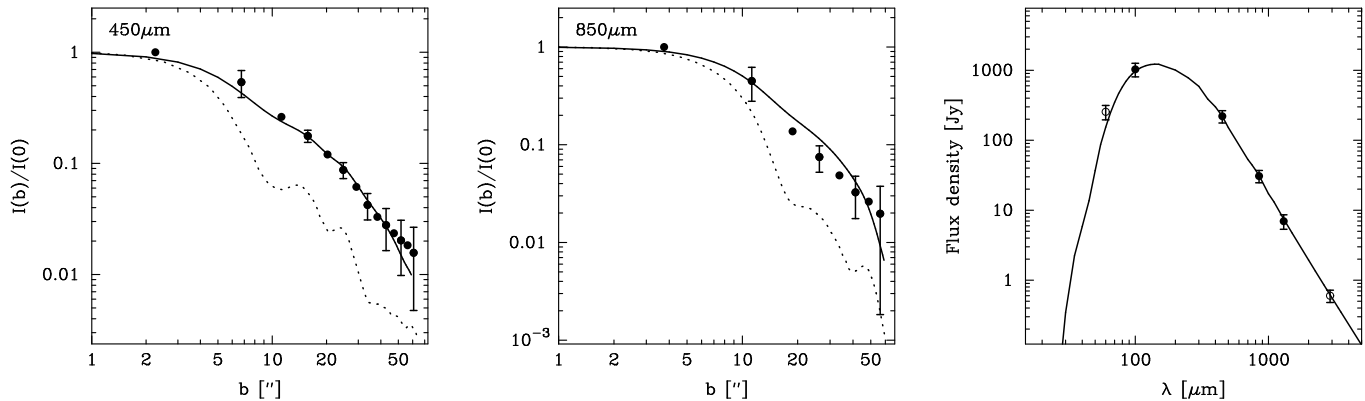}
 \caption{Best fit model, using a density structure described by a
 single power-law, compared with observed radial brightness
 distributions and the SED.  In the $\chi^2$-analysis of the radial
 brightness distributions only the data points separated by one full
 beam width were used, shown here with error bars.  The error bars
 represent the rms scatter of the observations within each of the
 bins and are a combination of the noise as well as gradients and
 departure from spherical symmetry in the brightness map.  The
 dotted line is the azimuthally averaged SCUBA beam at the time of
 the observation.  In the SED panel the observations, represented by
 circles with error bars (open circles were not used in the
 $\chi^2$-analysis), are overlayed with the output from the best fit
 model.} 
 \label{bestfit_dust} 
 \end{figure*}

The quality of the best fit model can be judged from the reduced $\chi^2$ obtained from
\begin{equation}
\chi^2_{\mathrm{red}} = \frac{\chi^2_{\mathrm{min}}}{N-p},
\end{equation} 
where $p$ is the number of adjustable parameters. It is found that the
best fit model presented above has $\chi^2_{\mathrm{red}}$$=$0.1 when
the SED is used as constraint and $\chi^2_{\mathrm{red}}$$=$0.4 for
the 450\,$\mu$m and 850\,$\mu$m radial brightness distributions,
respectively.  In all, the observations are well reproduced.
 The best fit solution is presented in Table~\ref{dusty}.
The
inner radius, $r_{\mathrm i}$, of the envelope is calculated to be
4.8$\times$10$^{14}$\,cm (32\,AU) fixing the outer radius to
1.2$\times$10$^{17}$\,cm (8000\,AU or 50$\arcsec$). \citet{Mundy90}
estimated the radial size of the envelope to be $\sim$4400\,AU from
observations of ammonia emission.  The best fit model is presented in
Fig.~\ref{bestfit_dust} overplotted on the observational constraints.
The total mass of molecular hydrogen contained within the outer radius of 50$\arcsec$ is estimated to be 
5.4\,M$_{\sun}$. 
\citet{Blake94} derived a H$_2$ mass content of
$\sim$0.5$-$0.75\,M$_{\sun}$ within a 10$\arcsec$ radius from an
excitation analysis of the observed C$^{17}$O molecular line emission
assuming its abundance to be 3.8$\times$10$^{-8}$. The present
analysis gives a mass of $\sim$0.7\,M$_{\sun}$ within the same radius,
in excellent agreement. Thus, \object{IRAS 16293--2422} appears to have
one of the most massive envelopes of the known class~0 protostars
\citep{Andre00, Jorgensen02}.

Although disk emission is typically only responsible for a small fraction of the
total flux at sub-millimetre wavelengths it can contribute to the fluxes of the innermost
points on the brightness profiles leading to a steeper inferred density profile.
Tests in which the flux within a radius of one beam was reduced by 50\%
 indicate that the best fit value of $\alpha$ is reduced by 0.1-0.2.

\begin{figure*} 
 \includegraphics[width=180mm]{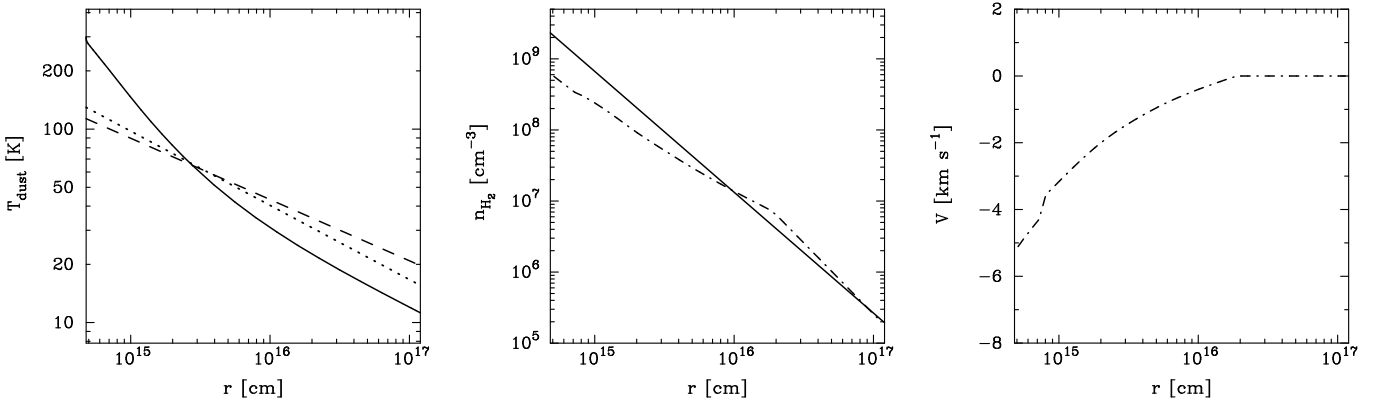}
 \caption{Properties of the circumstellar envelope around \object{IRAS 16293--2422} obtained from the 
 dust modelling. Shown are the dust temperature (left) and density (middle) structures 
 for best fit model (full  line) assuming a single power-law distribution of the density. 
 In the temperature panel, an optically thin prediction using an opacity
 law $\kappa_{\nu}$$\propto$$\nu^{\beta}$ with $\beta$$=$1 (dotted
 line; $T_{\mathrm d}$$\propto$$r^{-0.4}$) and $\beta$$=$2 (dashed
 line; $T_{\mathrm d}$$\propto$$r^{-0.33}$), is also shown. 
 In the density panel the best fit Shu-type collapsing envelope model is also shown (dash-dotted line). 
 The velocity structure obtained from the Shu infall model is shown in the right panel.}  
 \label{tdust}
\end{figure*}

The temperature and density structures obtained from the best fit model are
presented in Fig.~\ref{tdust}. For comparison, the predicted temperature
structure based upon an optically thin approximation
\citep{Chandler00} and scaled to the luminosity of \object{IRAS
16293--2422} is shown for two different opacity laws. Both
predict the dust temperature to follow a single power-law. 
Clearly, the temperature structure obtained from the detailed
radiative transfer analysis is not well described by a
power-law and has a significantly steeper gradient in the inner parts
of the envelope where optical depth effects are important.  
The interstellar radiation field is potentially important for the
temperature structure in the outer parts of the envelope. However,
detailed modelling (S.~Doty, priv. comm.)  shows, for the envelope
around \object{IRAS 16293--2422}, that this effect is small when
assuming a typical interstellar radiation field. The difference in
the dust temperature is $\sim$10-20\% (a few K) at
$r$$\gtrsim$8$\times$10$^{16}$\,cm. 

The dust opacities adopted in the previous analysis might not be
appropriate in the inner hot parts of the envelope where the ices
start to evaporate off the grains. Instead opacities for bare grains,
without any ice mantles, should preferably be used. DUSTY is not set
up to allow dust opacities with radial dependence so only the limiting
cases with or without ice mantles can be compared.
To test the sensitivity of the derived envelope parameters on the
adopted set of dust opacities, the analysis is
repeated using the bare grain opacities presented by
\citet{Ossenkopf94} (column 2 in their Table~1; hereafter OH2). From
the radial brightness distributions the same range of $\alpha$ and
$\Delta R$ as when using OH5 is obtained. The optical depth, however,
is significantly reduced by about a factor of two, and so is the total
mass of the envelope. 
Adopting the model parameters as derived in Table~\ref{dusty}, i.e.,
$\tau_{100}$$=$4.5, $\alpha$$=$1.7, and $\Delta R$$=$250, but using
the OH2 opacities increases the flux at
60\,$\mu$m by $\sim$20\% compared with the OH5 model. The 100\,$\mu$m
flux is not significantly affected whereas the 1.3\,mm flux is roughly
twice that of the OH5 flux.  A model with dust properties varying with
radius (OH2 in the inner warm parts and OH5 in the outer cool parts)
would thus serve to improve the fit to the SED.  Since the bulk of the
mass (99.8\%) is at low
temperatures where the grains are coated with ice mantles, the model
parameters obtained with the OH5 opacities were used in the chemical
analysis.

\begin{figure*} 
 \includegraphics[width=180mm]{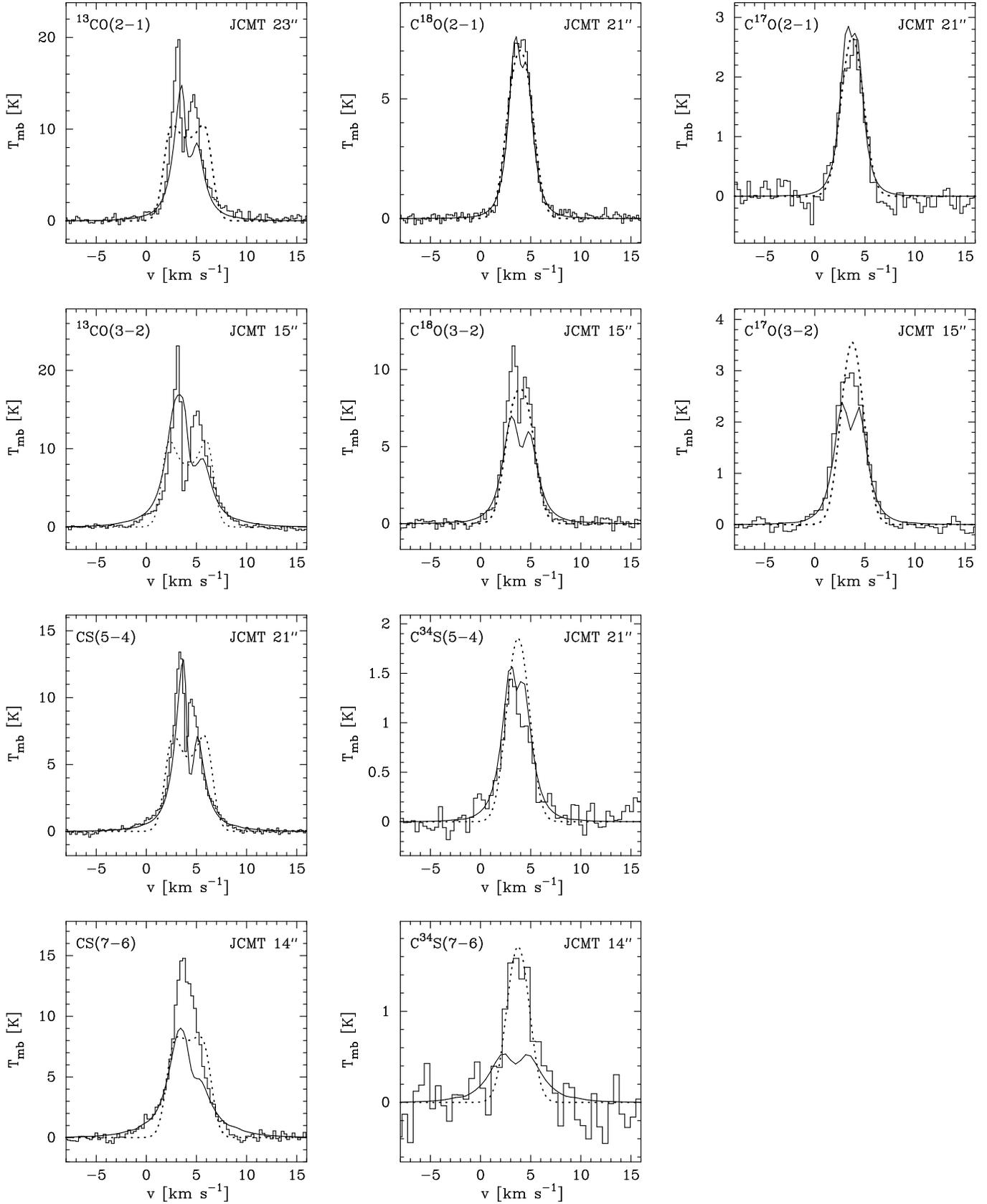}
 \caption{Best fit CO and CS models using a static envelope (dotted line)
 and a Shu-type collapsing model (full line), overlayed on the
 observed spectra (histograms) (see text for details). 
 The calibration uncertainty in the
 intensity scale for the observed spectra is $\sim$15\%.}  
 \label{overlay} 
\end{figure*}

 In addition to the analysis of the continuum emission the observed molecular
line emission is useful in constraining the physical properties of the envelope.
Traditionally, CO and CS line emission have been extensively used for this purpose and are
adopted here to test the validity of the best fit model obtained from the dust analysis.
Using the radiative transfer code presented in Sect.~\ref{gasmod}, the total CO and CS
abundances relative to H$_2$ obtained for the best fit model presented in Table~\ref{dusty}
are $\sim$4$\times$10$^{-5}$ and $\sim$3$\times$10$^{-9}$, respectively. 
In the $\chi^2$-analysis only the velocity-integrated intensities in the lines were used.
These values are within a factor of about two of what is commonly derived for YSOs
\citep{Tak00}. The abundances in combination with the quality of the fits 
(Fig.~\ref{overlay}; see also Sect.~\ref{molabund}),
in particular the ratios among various transitions which
are sensitive to the gas temperature and density, are reassuring and further strengthen the adopted
physical model.

In the modelling of the molecular line emission 
the gas temperature is assumed to follow that of the dust.
In models which self-consistently treat 
the energy balance the gas temperature is generally lower than that of the dust 
in the outer regions due to imperfect gas-grain coupling 
\citep{Ceccarelli96, Doty97, Ceccarelli00a}. 
To test the effects of a departure of the gas temperature from that of
the dust due to gas-grain decoupling in the outer regions,
the dust temperature was scaled by a constant factor. For
$T_{\mathrm{gas}}$$\lesssim$0.7$\times$$T_{\mathrm{dust}}$ the
envelope becomes too cool to fit the observed line intensity
ratios. Thus, the gas temperature appears to follow that of the dust
within $\sim$30\% in the region probed by the CO emission. 

 The static envelope model fails to explain the details of the
individual spectra and the potential of the molecular emission to constrain the velocity fields will 
be  investigated further in Sect.~\ref{infall}. 
The derived abundances are not very sensitive to the adopted value of $\alpha$, within the 
limits derived from the dust radiative transfer model. Similarly, increasing the outer radius by a factor
of two only marginally affects the line intensities.
The abundances derived
for a wide variety of molecular species are further presented in Sect.~\ref{molabund}.

Thus, the emerging picture from the analysis is that the envelope around
\object{IRAS 16293--2422} indeed has a region of dense and hot gas
inside a radius of $\sim$2$\times$10$^{15}$ cm (150 AU, 1$\arcsec$), 
with temperatures decreasing to $\sim$10\,K at $\sim$10$^{17}$\,cm
(8000\,AU).

\begin{figure*} 
 \includegraphics[width=180mm]{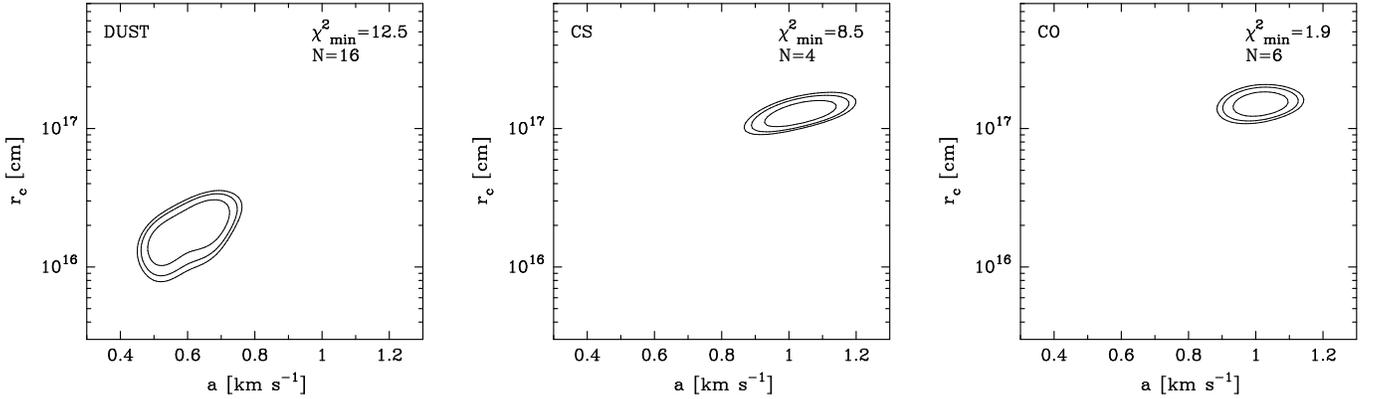}
 \caption{$\chi^2$-maps showing the sensitivity of the Shu-collapse
 model to the adjustable parameters in the modeling of the dust
 continuum emission (left) and the CS (CS and C$^{34}$S transitions;
 centre) and CO ($^{13}$CO, C$^{17}$O, and C$^{18}$O transitions;
 right) line emission compared to observations.  Contours are drawn
 at $\chi^2_{\mathrm{min}}$$+$(2.3, 4.6, 6.2) indicating the 68\%
 (`1$\sigma$'), 90\%, and 95\% (`2$\sigma$') confidence levels,
 respectively. The number of observational constraints used, $N$,
 are also shown.  The quality of the best fit model can be estimated
 from the reduced chi-squared statistic
 $\chi^{2}_{\mathrm{red}}$$=$$\chi^{2}_{\mathrm{min}}$/($N$$-$2).}
 \label{shu_chi2_fig} 
\end{figure*}

\subsection{Infall model}
\label{infall}
Current theories of star formation state that protostars are formed
from the gravitational collapse of cloud cores consisting of gas and
dust \citep[e.g.,][]{Shu87}.  There is now growing evidence that
low-mass protostars have parts of their circumstellar material in a
state of collapse \citep[e.g.,][]{Myers00} and an obvious extension to
the previous analysis, which uses a static envelope with a density
distribution described by a single power-law, is to attempt to reproduce
the results with an infall
model.  That the circumstellar envelope around \object{IRAS
16293--2422} is in a state of collapse has been inferred previously
from modelling of millimetre CS observations \citep{Walker86, Zhou95,
Narayanan98}, although such conclusions based on low spatial resolution
data have been questioned by \citet{Menten87}.
Furthermore, Ceccarelli et~al.\ (2000a,b) \nocite{Ceccarelli00a}
\nocite{Ceccarelli00b} have suggested an infall model based upon a
physical-chemical model.  Here, the observed continuum and molecular
line emission will be analyzed using the well known collapse model
presented by \citet{Shu77}.

In the Shu inside-out collapse model the self-similar solution is
presented in terms of the dimensionless variable $x$$=$$r/at$, where $r$
is the radial distance scale, and characterized by the isothermal
speed of sound, $a$, and the time after onset of collapse, $t$. The
location of the collapsing wave front at any instant $t$ is described
by $r_{\mathrm c}$$=$$at$. The density $\rho$ and velocity $u$ have
the form
\begin{equation}
\label{shu_dens}
\rho(r,t) = \frac{\alpha(x)}{4\pi Gt^2},\ u(r,t) = a v(x),
\end{equation}
where $G$ is the gravitational constant and $\alpha(x)$ and $v(x)$ are tabulated by \citet{Shu77}.
In the static part of the envelope ($x$$>$1) 
\begin{equation}
\alpha(x)=\frac{2}{x^2}, \ v(x)=0.
\end{equation}
The asymptotic behavior as $x$$\rightarrow$0 of the solution is
\begin{equation}
\alpha(x)=\left( \frac{m_0}{2x^3} \right)^{1/2},\ v(x)=-\left (\frac{2m_0}{x} \right)^{1/2},
\end{equation}
where $m_0$ is equal to 0.975 for this particular solution.  Given the
limited resolution provided by the JCMT at 450 and 850 microns it is
not surprising that the observations are well described by a single
power-law with $\alpha$ in the range $\sim$1.5-2.0 which are the two
extremes obtained from the Shu-model.
The estimated value of $\alpha$ will
depend on the location of the collapsing wavefront, $r_{\mathrm c}$.
In comparison, \citet{Jorgensen02} derived values of $\alpha$ 
for a large number of class~0 protostars and found values 
typically in the range $\sim$1.3-2.0 .

The input parameters to DUSTY are the same as for the single power-law
models (see Table~\ref{dusty}). In addition, the envelope size was
fixed to a radius 5000\,AU. Making the envelope larger will produce
increasingly worse fits to the radial brightness distributions obtained
from the SCUBA observations. The sensitivity of the two adjustable
parameters $a$ and $r_{\mathrm c}$$=$$at$ in the modelling is shown in
Fig.~\ref{shu_chi2_fig} (left panel)
where the observational constraints used are the
SED and the SCUBA 450\,$\mu$m radial brightness distribution.  The
best fit model is obtained using $a$$\sim$0.6\,km\,s$^{-1}$ and
$r_{\mathrm c}$$\sim$2$\times$10$^{16}$\,cm, putting the age at
$\sim$10$^4$\,yr.

The mass accretion rate can be estimated from
\begin{equation}
\dot{M}=\frac{m_0a^3}{G},
\end{equation}
where $G$ is the gravitational constant.  For the best fit model
$\dot{M}$$\sim$5$\times$10$^{-5}$\,M$_{\sun}$\,yr$^{-1}$ is obtained.
Such a high accretion rate is able to account for the source luminosity.
The total mass contained in the envelope within 5000\,AU is
$\sim$3.3\,M$_{\sun}$, consistent with the estimate obtained from the
static envelope model within the same radius.

The relative success of the dust modelling using a static envelope,
with a single power-law to describe the density structure, makes it
hard to discriminate between the two models in the present
analysis. In Fig.~\ref{tdust} the density and velocity structures
obtained from the best fit Shu model are presented and, for the 
density, compared with results from the static envelope model. 
The largest discrepancy
occurs at small radial distances where the collapsing envelope model
predicts about a factor of two to three lower densities. We stress 
the observational data set used in the dust modelling is not directly probing this region. 
At larger radii the model is better constrained and the two models agree well.
It should be noted that the best fit single power-law model
gives slightly better reduced $\chi^2$ values for the combined set of
observations.  However, the molecular data provide further constraints
since they have the potential to probe the large scale velocity field.

In Fig.~\ref{overlay}, spectra of CO and CS line emission as observed with the JCMT are presented.
Lines which are optically thick, like CS, $^{13}$CO and C$^{18}$O ($J$$=$3$\rightarrow$2), 
show a distinct, narrow, absorption feature near the stellar velocity. This feature is due to effective 
self-absorption in the outer cool parts of the envelope. 
Also, the degree of the asymmetry in the line profiles 
increases with the optical depth in the lines. In the optically thin lines the self-absorption feature 
disappears and the lines are well described by a single gaussian profile.
The width of the self-absorption 
feature constrains the turbulent velocity to $\sim$0.3\,km\,s$^{-1}$ in the outer envelope. For
simplicity this value is adopted throughout the envelope. A turbulent velocity component
varying with radius is beyond the scope of this article, 
see however Stark et~al. (2002, in prep.).

The observed CS emission (including that from C$^{34}$S) is analyzed
using the abundances derived from the static envelope model
 presented in Sect.~\ref{dust_oh5} (see also Sect.~\ref{molabund}). 
From only the integrated intensities it is possible to
constrain both $a$ and $r_{\mathrm c}$ as is shown in
Fig.~\ref{shu_chi2_fig} (middle panel).  
The model spectra are presented in Fig.~\ref{overlay} together with
the observations. The fit to the integrated intensities is worse than that
obtained from the single power-law model  presented in
Sect.~\ref{dust_oh5}, however. In particular the
C$^{34}$S($J$$=$7$\rightarrow$6) line is poorly reproduced in the
infall model.  
Using also the line profiles as constraints we find
that a model where the collapsing wavefront is located at
$\sim$1.0$\times$10$^{17}$\,cm and the value of $a$ is
$\sim$0.9\,km\,s$^{-1}$ best reproduces the observations.

Analyzing the CO emission ($^{13}$CO, C$^{18}$O, and C$^{17}$O) gives
yet another set of estimates. At first the abundances derived from the static envelope 
models presented in  Sect.~\ref{dust_oh5} (see also Sect.~\ref{molabund}) are adopted.
From the analysis of the integrated intensities shown in Fig.~\ref{shu_chi2_fig} (right panel) 
and the line profiles presented in Fig.~\ref{overlay}
a Shu-model with $a$$\sim$1.0\,km\,s$^{-1}$ and 
$r_{\mathrm c}$$\sim$1.2$\times$10$^{17}$\,cm reproduces the CO observations well, 
in excellent agreement with the CS modelling. 
The derived dynamical age of the system is $\sim$3$\times$10$^{4}$\,yr.
In contrast with the CS modelling, the fit to the integrated CO intensities 
is equally good as obtained for the single power-law model.
As discussed in Sect.~\ref{molabund}, the CO abundance obtained from the static envelope 
model is about a factor of 2$-$3 lower than what is typically observed for interstellar gas. 
If instead the CO abundance relative to H$_2$ is
assumed to be the `standard' interstellar value of 1$\times$10$^{-4}$
and the standard isotopic ratios are assumed ( [CO/$^{13}$CO]=60, [CO/C$^{17}$O]=2500, and
[C$^{18}$O/C$^{17}$O]=3.9) the estimate
of $a$ is $\sim$0.75 and $r_{\mathrm c}$$\sim$3$\times$10$^{17}$\,cm. However, the quality of the fit becomes worse in this case.

The results obtained from the best fit Shu-model are presented in Fig.~\ref{overlay} 
overlayed onto the observed CO and CS spectra. The integrated intensities and, to some extent, line profiles
can be modelled with the spherically symmetric infall solution. 
The details of the spectra will, however, intricately depend on 
the adopted geometry, velocity fields, and chemical gradients.
The
velocity field, in particular position-velocity maps of the source, form a 
stringent test of the dynamical models.  For \object{IRAS 16293--2422} 
it will likely be necessary to include a rotational
component to the velocity field \citep{Menten87, Zhou95, Narayanan98} .

Other estimates of the infall radius, based on analysis of molecular
line emission, range between about 5 and 15$\times$10$^{16}$\,cm
\citep{Walker86, Zhou95, Narayanan98, Ceccarelli00a} in excellent
agreement with the values obtained here from analysis of CO and CS
emission. However, there appears to be a discrepancy between the dust
and molecular line analysis, possibly reflecting the fact that the
simple Shu-collapsing core model is not fully adequate to describe the
state of the infalling material and/or that some of the CO and CS
emission is associated with the outflow and surrounding
cloud. Moreover, gradients in the molecular abundances will affect the
parameters derived.

\begin{table*}
   \caption[]{Derived abundances using a constant molecular abundance 
      $f_X$ relative to H$_2$ throughout the envelope.}
      \label{abundances} 
      $$
      \begin{array}{p{0.1\linewidth}p{0.2\linewidth}ccccccccccc}
	 \hline
	 \noalign{\smallskip}
     	  & 
	  & 
     	  &
     	  &
     	  &
     	  &
     	  &
     	  &
     	  &
     	  &
     	 \multicolumn{1}{c}{{E_{\mathrm{min}}}^{\mathrm d}} &
     	 &
     	 \multicolumn{1}{c}{{E_{\mathrm{max}}}^{\mathrm e}}\\
     	 
     	 \noalign{\smallskip}
     	 \multicolumn{2}{c}{\mathrm{Molecule}} & 
	 \multicolumn{1}{c}{f_X} & &
     	 \multicolumn{1}{c}{{\chi^2_{\mathrm{red}}}^{\mathrm a}} & &
     	 \multicolumn{1}{c}{{N}^{\mathrm b}} & &
     	 \multicolumn{1}{c}{{f_X^{\mathrm{old}}}^{\mathrm c}} & &
     	 \multicolumn{1}{c}{\mathrm{[K]}} & &
     	 \multicolumn{1}{c}{\mathrm{[K]}}\\
     	 
	 \noalign{\smallskip}
	 \hline
	 \noalign{\smallskip}
     	 $^{13}$CO	   & carbon monoxide	 & \phantom{\mathrm{<}}6.5\times 10^{-7}\phantom{^0\dagger}  && \phantom{0}0.2 && \phantom{0}3  && \phantom{\mathrm{<}}1.6\times 10^{-6}{\dagger}\phantom{^0}		&& \phantom{0}15.9   && \phantom{0}31.7  \\
     	 C$^{18}$O	   &			 & \phantom{\mathrm{<}}6.2\times 10^{-8}\phantom{^0\dagger}  && \phantom{0}0.5 && \phantom{0}2  && \phantom{\mathrm{<}}1.0\times 10^{-7}\phantom{^0\dagger}             && \phantom{0}15.8   && \phantom{0}31.6  \\
	 C$^{17}$O	   &			 & \phantom{\mathrm{<}}1.6\times 10^{-8}\phantom{^0\dagger}  && \phantom{0}0.1 && \phantom{0}2  && \phantom{\mathrm{<}}3.8\times 10^{-8}\phantom{^0\dagger}             && \phantom{0}16.2   && \phantom{0}32.4  \\
     	 HCO$^{+}$	   & formyl ion 	 & \phantom{\mathrm{<}}1.4\times 10^{-9}{\dagger}\phantom{^0}	   &&		  10.9 && \phantom{0}3 && \phantom{\mathrm{<}}1.8\times 10^{-9}{\dagger}\phantom{^0}    && \phantom{0}25.7   && \phantom{0}42.8  \\
     	 H$^{13}$CO$^{+}$  &			 & \phantom{\mathrm{<}}2.4\times 10^{-11}\phantom{\dagger} && \phantom{0}0.1 && \phantom{0}3 && \phantom{\mathrm{<}}7.5\times 10^{-12}\phantom{\dagger}                 && \phantom{0}25.0   && \phantom{0}41.6  \\
     	 HC$^{18}$O$^{+}$  &			 & \phantom{\mathrm{<}}6.4\times 10^{-12}\phantom{\dagger} &&	       \cdots	&& \phantom{0}1 && \phantom{\mathrm{<}}3.5\times 10^{-12}\phantom{\dagger}              && \phantom{0}40.9   && \phantom{0}40.9  \\ 
    	 DCO$^{+}$	   &			 & \phantom{\mathrm{<}}1.3\times 10^{-11}\phantom{\dagger} && \phantom{0}0.3 && \phantom{0}2 && \phantom{\mathrm{<}}1.5\times 10^{-11}\phantom{\dagger}                 && \phantom{0}20.7   && \phantom{0}51.9  \\

     	 CN		   & cyanogen		 & \phantom{\mathrm{<}}8.0\times 10^{-11}\phantom{\dagger} && \phantom{0}1.2 && \phantom{0}4 && \phantom{\mathrm{<}}1.0\times 10^{-10}\phantom{\dagger}                 && \phantom{0}16.3   && \phantom{0}32.7  \\
     	 HCN		   & hydrogen cyanide	 & \phantom{\mathrm{<}}1.1\times 10^{-9}{\dagger}\phantom{^0}	     && \phantom{0}2.3 && \phantom{0}3 && \phantom{\mathrm{<}}1.9\times 10^{-9}{\dagger}\phantom{^0}    && \phantom{0}25.5   && \phantom{0}42.5  \\
     	 H$^{13}$CN	   &			 & \phantom{\mathrm{<}}1.8\times 10^{-11}\phantom{\dagger} && \phantom{0}6.2 && \phantom{0}2 && \phantom{\mathrm{<}}1.4\times 10^{-11}\phantom{\dagger}                 && \phantom{0}24.9   && \phantom{0}41.4  \\
     	 HC$^{15}$N	   &			 & \phantom{\mathrm{<}}7.6\times 10^{-12}\phantom{\dagger} &&	       \cdots	&& \phantom{0}1 && \phantom{\mathrm{<}}7.0\times 10^{-12}\phantom{\dagger}              && \phantom{0}24.8   && \phantom{0}24.8  \\
     	 DCN		   &			 & \phantom{\mathrm{<}}1.3\times 10^{-11}\phantom{\dagger} && \phantom{0}3.5 && \phantom{0}2 && \phantom{\mathrm{<}}2.5\times 10^{-11}\phantom{\dagger}                 && \phantom{0}20.9   && \phantom{0}52.1  \\
     	 HNC		   & hydrogen isocyanide & \phantom{\mathrm{<}}6.9\times 10^{-11}\phantom{\dagger} && \phantom{0}0.2 && \phantom{0}3 && \phantom{\mathrm{<}}1.5\times 10^{-10}\phantom{\dagger}                 && \phantom{0}26.1   && \phantom{0}43.5  \\
	 HN$^{13}$C	   &			 & \phantom{\mathrm{<}}5.2\times 10^{-12}\phantom{\dagger} &&	       \cdots	&& \phantom{0}1 && \phantom{\mathrm{<}}2.5\times 10^{-12}\phantom{\dagger}              && \phantom{0}25.1   && \phantom{0}25.1  \\
     	 DNC		   &			 & \phantom{\mathrm{<}}4.2\times 10^{-12}\phantom{\dagger} && \phantom{0}3.0 && \phantom{0}2 && \phantom{\mathrm{<}}5.0\times 10^{-12}\phantom{\dagger}                 && \phantom{0}22.0   && \phantom{0}36.6  \\
     	 HC$_3$N	   & cyanoacetylene	 & \phantom{\mathrm{<}}1.5\times 10^{-10}\phantom{\dagger} && \phantom{0}1.8 && \phantom{0}3 && \phantom{\mathrm{<}}2.5\times 10^{-11}\phantom{\dagger}                 &&	     131.0   && 	  177.3  \\
     	 CH$_3$CN	   & methyl cyanide	 & \phantom{\mathrm{<}}1.4\times 10^{-10}\phantom{\dagger} && \phantom{0}5.0 && \phantom{0}7 && \phantom{\mathrm{<}}1.5\times 10^{-10}\phantom{\dagger}                 && \phantom{0}68.9   && 	  257.9  \\
     	 HNCO		   & isocyanic acid	 & \phantom{\mathrm{<}}1.3\times 10^{-10}\phantom{\dagger} && \phantom{0}1.0 && \phantom{0}2 && \phantom{\mathrm{<}}1.7\times 10^{-10}\phantom{\dagger}                 &&	     112.6   && 	  126.6  \\

     	 C$_2$H 	   & ethynyl		 & \phantom{\mathrm{<}}2.1\times 10^{-10}\phantom{\dagger} && \phantom{0}0.3 && \phantom{0}4 && \phantom{\mathrm{<}}2.5\times 10^{-10}\phantom{\dagger}                 && \phantom{0}25.1   && \phantom{0}62.9  \\
     	 C$_2$D 	   &			 & \phantom{\mathrm{<}}3.5\times 10^{-11}\phantom{\dagger} && \phantom{0}0.0 && \phantom{0}2 && \phantom{\mathrm{<}}4.5\times 10^{-11}\phantom{\dagger}                 && \phantom{0}20.7   && \phantom{0}20.7  \\ 
     	 C$_3$H$_2$	   & cyclopropenylidene  & \phantom{\mathrm{<}}1.6\times 10^{-11}\phantom{\dagger} && \phantom{0}4.1 && \phantom{0}6 && \phantom{\mathrm{<}}3.5\times 10^{-11}\phantom{\dagger}                 && \phantom{0}19.5   && \phantom{0}86.9  \\
     	 CH$_3$C$_2$H	   & methyl acetylene	 & \phantom{\mathrm{<}}1.4\times 10^{-9}\phantom{^0\dagger}  && \phantom{0}2.7 && \phantom{0}6  && \phantom{\mathrm{<}}6.5\times 10^{-10}\phantom{\dagger}              && \phantom{0}74.6   && 	  163.2  \\

     	 o-H$_2$CO	   & formaldehyde	 & \phantom{\mathrm{<}}6.9\times 10^{-10}\phantom{\dagger} && \phantom{0}4.1 && \phantom{0}9 && \phantom{\mathrm{<}}5.2\times 10^{-10}\phantom{\dagger}                 && \phantom{0}21.9   && 	  174.0  \\
     	 p-H$_2$CO	   &			 & \phantom{\mathrm{<}}5.1\times 10^{-10}\phantom{\dagger} && \phantom{0}2.9 && \phantom{0}8 && \phantom{\mathrm{<}}1.8\times 10^{-10}\phantom{\dagger}                 && \phantom{0}21.0   && 	  240.7  \\
     	 o-H$_2$$^{13}$CO  &			 & \phantom{\mathrm{<}}1.2\times 10^{-11}\phantom{\dagger} && \phantom{0}3.7 && \phantom{0}4 &&      \cdots	                                                        && \phantom{0}21.7   && \phantom{0}61.3  \\
     	 p-H$_2$$^{13}$CO  &			 & \phantom{\mathrm{<}}9.5\times 10^{-12}\phantom{\dagger} && \phantom{0}3.6 && \phantom{0}2 &&      \cdots	                                                        && \phantom{0}51.1   && \phantom{0}98.4  \\
     	 o-HDCO 	   &			 & \phantom{\mathrm{<}}1.0\times 10^{-10}\phantom{\dagger} && \phantom{0}0.1 && \phantom{0}2 && \phantom{\mathrm{<}}7.3\times 10^{-11}\phantom{\dagger}                 && \phantom{0}52.4   && \phantom{0}56.3  \\
     	 p-HDCO 	   &			 & \phantom{\mathrm{<}}5.1\times 10^{-11}\phantom{\dagger} && \phantom{0}5.0 && \phantom{0}2 && \phantom{\mathrm{<}}2.5\times 10^{-11}\phantom{\dagger}                 && \phantom{0}30.8   && \phantom{0}62.7  \\
     	 CH$_3$OH	   & methanol		 & \phantom{\mathrm{<}}1.7\times 10^{-9}\phantom{^0\dagger}  && \phantom{0}5.7 &&	   23  && \phantom{\mathrm{<}}4.4\times 10^{-9}\phantom{^0\dagger}              && \phantom{0}15.5   && 	  187.6  \\	 
	 CH$_2$CO	   & ketene		 & \phantom{\mathrm{<}}5.3\times 10^{-10}\phantom{\dagger} && \phantom{0}9.2 && \phantom{0}2 && \phantom{\mathrm{<}}1.8\times 10^{-10}\phantom{\dagger}                 && \phantom{0}88.0   && 	  160.0  \\    
     	 HCOOH  	   & formic acid	 & \mathrm{<}3.0\times 10^{-10}\phantom{\dagger} && \cdots	  && \cdots	  && \mathrm{<}3.0\times 10^{-10}\phantom{\dagger}                                      &&	    \cdots   && 	\cdots   \\
     	 CH$_3$CHO	   & acetaldehyde	 & \mathrm{<}6.0\times 10^{-11}\phantom{\dagger} && \cdots	  && \cdots	  && \mathrm{<}1.0\times 10^{-10}\phantom{\dagger}                                      &&	    \cdots   && 	\cdots   \\
     	 CH$_3$OCH$_3$     & dimethyl ether	 & \mathrm{<}6.0\times 10^{-10}\phantom{\dagger} && \cdots	  && \cdots	  && \mathrm{<}2.0\times 10^{-9}\phantom{^0\dagger}                                     &&	    \cdots   && 	\cdots   \\
     	 HCOOCH$_3$	   & methyl formate	 & \mathrm{<}5.0\times 10^{-9}\phantom{^0\dagger}  && \cdots	   && \cdots	                              &&		   \cdots                               &&	    \cdots   && 	\cdots   \\	 

     	 CS		   & carbon monosulfide  & \phantom{\mathrm{<}}3.0\times 10^{-9}\phantom{^0\dagger}  && \phantom{0}0.5 && \phantom{0}3  && \phantom{\mathrm{<}}1.1\times 10^{-9}\phantom{^0\dagger}             && \phantom{0}35.3   && \phantom{0}65.8  \\
     	 C$^{34}$S	   &			 & \phantom{\mathrm{<}}1.2\times 10^{-10}\phantom{\dagger} && \phantom{0}0.9 && \phantom{0}3  && \phantom{\mathrm{<}}5.0\times 10^{-11}\phantom{\dagger}                && \phantom{0}34.7   && \phantom{0}64.8  \\

     	 SO		   & sulfur monoxide	 & \phantom{\mathrm{<}}4.4\times 10^{-9}\phantom{^0\dagger}  && \phantom{0}2.4 && \phantom{0}9  && \phantom{\mathrm{<}}3.9\times 10^{-9}\phantom{^0\dagger}             && \phantom{0}35.0   && \phantom{0}87.5  \\
     	 SO$_2$ 	   & sulfur dioxide	 & \phantom{\mathrm{<}}6.2\times 10^{-10}\phantom{\dagger} && \phantom{0}6.1 && 	  10 && \phantom{\mathrm{<}}1.5\times 10^{-9}\phantom{^0\dagger}                && \phantom{0}19.2   && \phantom{0}82.5  \\
     	 OCS		   & carbonyl sulfide	 & \phantom{\mathrm{<}}7.0\times 10^{-9}\phantom{^0\dagger}  && \phantom{0}6.5 && \phantom{0}2  && \phantom{\mathrm{<}}7.1\times 10^{-9}\phantom{^0\dagger}             &&	     122.6   && 	  237.0  \\
     	 HCS$^{+}$	   & thioformyl ion	 & \phantom{\mathrm{<}}2.4\times 10^{-11}\phantom{\dagger} && \phantom{0}7.0 && \phantom{0}2 && \phantom{\mathrm{<}}2.0\times 10^{-11}\phantom{\dagger}                 && \phantom{0}30.7   && \phantom{0}73.7  \\
     	 O$^{13}$CS	   &			 & \phantom{\mathrm{<}}6.7\times 10^{-10}\phantom{\dagger} && \phantom{0}5.6 && \phantom{0}2  && \phantom{\mathrm{<}}2.8\times 10^{-10}\phantom{\dagger}                && \phantom{0}99.5   && 	  122.2  \\
     	 OC$^{34}$S	   &			 & \phantom{\mathrm{<}}1.6\times 10^{-9}\phantom{^0\dagger}  && \phantom{0}1.8 && \phantom{0}3  && \phantom{\mathrm{<}}3.2\times 10^{-10}\phantom{\dagger}              && \phantom{0}99.8   && 	  147.7  \\
     	 H$_2$S 	   & hydrogen sulfide	 & \phantom{\mathrm{<}}1.6\times 10^{-9}\phantom{^0\dagger}  &&  	 \cdots && \phantom{0}1  && \phantom{\mathrm{<}}1.5\times10^{-9}\phantom{^0\dagger}             && \phantom{0}84.0   && \phantom{0}84.0  \\
     	 o-H$_2$CS	   & thioformaldehyde	 & \phantom{\mathrm{<}}2.0\times 10^{-10}\phantom{\dagger} && \phantom{0}8.8 && \phantom{0}6 && \phantom{\mathrm{<}}1.1\times 10^{-10}\phantom{\dagger}                 && \phantom{0}58.6   && 	  209.1  \\
     	 p-H$_2$CS	   &			 & \phantom{\mathrm{<}}1.8\times 10^{-10}\phantom{\dagger} && \phantom{0}2.9 && \phantom{0}4 && \phantom{\mathrm{<}}5.7\times 10^{-11}\phantom{\dagger}                 && \phantom{0}46.1   && \phantom{0}98.8  \\

     	 SiO		   & silicon monoxide	 & \phantom{\mathrm{<}}5.9\times 10^{-11}\phantom{\dagger} && \phantom{0}2.2 && \phantom{0}4 && \phantom{\mathrm{<}}1.0\times 10^{-10}\phantom{\dagger}                 && \phantom{0}31.3   && \phantom{0}75.0  \\
     	 $^{29}$SiO	   &			 & \phantom{\mathrm{<}}8.8\times 10^{-12}\phantom{\dagger} && \phantom{0}5.5 && \phantom{0}3 && \phantom{\mathrm{<}}5.0\times 10^{-12}\phantom{\dagger}                 && \phantom{0}31.0   && \phantom{0}74.3  \\

	 \noalign{\smallskip}
	 \hline
      \end{array}
      $$
      \smallskip

      \noindent
      $^{\mathrm a}$ {The reduced $\chi^2$ of the best fit model. 
     		     Based on the $\chi^2$-analysis the derived abundances are accurate to about 20$-$30\% (1$\sigma$)
     		     when the fit is good, i.e., $\chi^2_{\mathrm{red}}$$\sim$1.}\\
      \noindent
      $^{\mathrm b}$ The number $N$ of independent observational constraints used in the modelling.\\
      \noindent
      $^{\mathrm c}$ Abundance derived by \citet{Blake94} and \citet{Dishoeck95} using a simple excitation analysis. \\
      \noindent
      $^{\mathrm d}$ Lowest energy of the upper level involved in the transitions used as constraints in the modelling. \\
      \noindent
      $^{\mathrm e}$ Highest energy of the upper level involved in the transitions used as constraints in the modelling.\\
      \noindent
      ${\dagger}$ Derived from optically thin isotope(s) assuming standard isotopic ratio(s).
\end{table*}

\section{Molecular abundances}
\label{molabund}

The basic envelope parameters derived from the dust radiative transfer
modelling performed in Sect.~\ref{dustres}, in particular the density
and temperature distributions, are used as input for the Monte Carlo
modelling of the molecular line emission. The static power-law model
is adopted; the abundances obtained with the best fitting infall model
generally differ by no more than $\sim$25\% for a constant abundance
model. However, models where a drastic enhancement in the abundance is
introduced (`jump-models') require 2$-$3 times larger abundances in
the inner hot part for a Shu-type collapsing model, reflecting the
significantly lower density compared to the static power-law model in
this region (Fig.~\ref{tdust}).
 Changing the envelope parameters describing the static power-law model
within the accepted range of values (Fig.~\ref{chi2maps}) typically affects the 
abundances obtained for the best fit model by less than $\pm$25\% for constant abundance models.
In `jump-models' the effect on lines which are sensitive to the conditions in the innermost dense and hot
regions can be higher, up to $\pm$50\%.

\subsection{Constant abundance models}
The abundances are initially assumed to be constant throughout the
envelope. For simplicity the observed lines are assumed to be broadened 
(in addition to thermal line broadening) by microturbulent motions only.  
The microturbulent velocity is set equal to 2\,km\,s$^{-1}$ throughout the envelope 
typically producing lines $\sim$4\,km\,s$^{-1}$ (FWHM) wide 
(see Sect.~\ref{discussion}). 
The presence of a global velocity field, e.g., infall, outflow,
or rotation, would
serve to reduce the optical depths of the line emission, so that the
abundances presented in Table~\ref{abundances} are strictly lower
limits.  However, such effects will not significantly increase the
inferred abundances since they are largely derived from optically thin
lines.  The observational constraints used in the modelling are the
total velocity-integrated line intensities.

In many cases the observations are well reproduced assuming a constant
abundance throughout the envelope, as seen from their reduced
$\chi^2$$\sim$1. 
In the cases where the fits are good the derived
abundances are generally consistent with typical values found in
quiescent molecular clouds. In addition, the isotopic ratios of
$^{18}$O/$^{17}$O$\sim$3.9 determined from CO observations and
$^{32}$S/$^{34}$S$\sim$25 from CS observations, agree well with
interstellar values \citep{Wilson94}.
A notable exception is the relatively low abundance, $\sim$7$\times$10$^{-11}$,
derived for HNC. The abundances derived by \citet{Blake94} and
\citet{Dishoeck95} agree surprisingly well with the new, more
accurate, estimates presented here (Table~\ref{abundances}), typically
within a factor of $\sim$2. Those abundances were derived from
statistical equilibrium equations assuming a constant temperature and
density. The agreement indicates that the adopted values were
representative of the region from which most of the submillimetre
emission arises.

 For the main isotopes of HCN and HCO$^+$ the emission is highly optically thick and the models are
relatively insensitive to the molecular abundance. The abundances of these molecules were instead 
estimated from the rarer isotopomers assuming a standard isotope ratio, i.e., $^{12}$C/$^{13}$C$=$60. 
The abundances obtained in this way fail to account for all of the observed flux 
in the HCN and HCO$^+$ lines, as evidenced by their high reduced $\chi^2$-values in Table~\ref{abundances}.
Adopting the best fit Shu-infall model derived from CO and CS observations and presented in Sect.~\ref{infall}
introduces a large-scale velocity field which reduces the line optical depths and increases the line intensities
thus improving the fit to observations. Material in the outflow can also contribute to these lines.

The observed transitions are only in LTE throughout the envelope for 
abundant molecules like CO and OCS, including their isotopomers observed here. 
For less abundant species, where collisional excitation is less efficient,
departures from LTE are found. For example, the level populations  of 
common molecules like CS and H$_2$CO are in LTE out to 
$\sim$1$-$2$\times$10$^{16}$\,cm.
For most molecules, populations of the observed lines
are in LTE within $\sim$2$\times$10$^{15}$\,cm.

From Table~\ref{abundances} it is also evident that the line emission
from several molecular species is not fitted well, in particular
molecules where the emission probes a large radial range, e.g., in the
case of H$_2$CO and CH$_3$OH.  In addition, the isotopic ratios
derived in many of these cases are far from their interstellar values
and what is commonly derived for these kind of objects. Typically, the
model intensities from lines sampling the inner parts of the envelope
are too low compared to observed values, whereas the opposite is true
for the lines probing the outer part of the envelope.  An obvious
explanation is that a steep gradient is present in the
abundances of these molecules.

\subsection{Jump models}
\label{jump_sect}
\begin{figure*} 
 \centering
 \includegraphics[width=160mm]{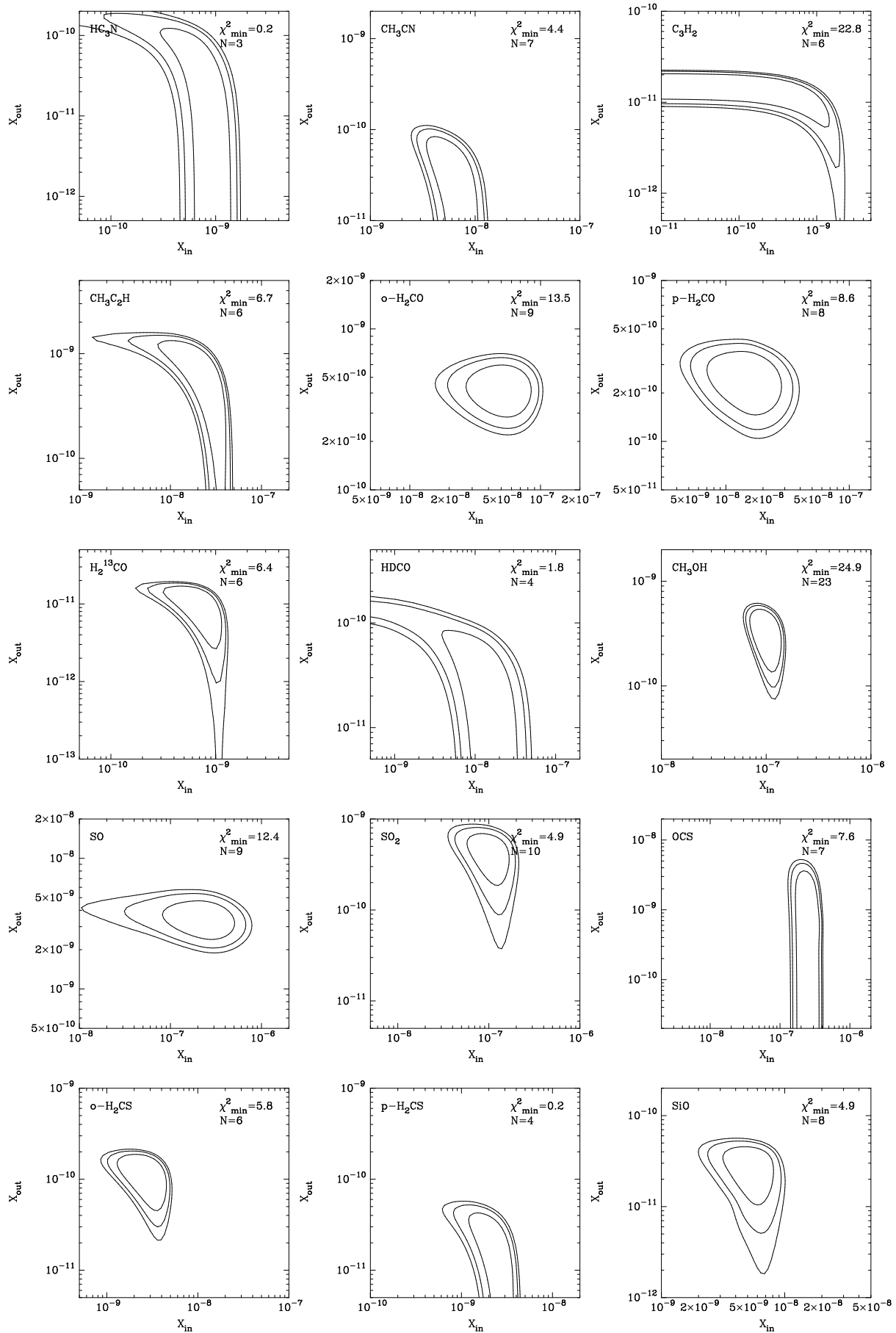}
 \caption{$\chi^2$-maps of various molecular species introducing a
 step function in abundance at $T_{\mathrm{gas}}$=90\,K.  Contours
 are drawn at $\chi^2_{\mathrm{min}}$$+$(2.3, 4.6, 6.2) indicating
 the 68\%, 90\%, and 95\% confidence levels, respectively. The
 number of observational constraints used, $N$, are also shown.  The
 quality of the best fit model can be estimated from the reduced
 chi-squared
 $\chi^{2}_{\mathrm{red}}$$=$$\chi^{2}_{\mathrm{min}}$/($N$$-$2).
 Fixed isotope ratios of $^{28}$SiO/$^{29}$SiO=20, OCS/OC$^{34}$S=20
 and OCS/O$^{13}$CS=60 were assumed.}  
 \label{chi2_jump}
\end{figure*}

In order to improve the quality of the fits, models with a jump in the
molecular abundances are considered. A jump is 
introduced at the radius in the envelope
where the temperature reaches 90\,K, at which point the ice starts to
evaporate from the grain mantles. In our models, this occurs at 
$\sim$2$\times$10$^{15}$\,cm or 150\,AU.  The free parameters in the modelling
are then the fractional abundances in the inner ($f_{\mathrm{in}}$;
$T$$>$90\,K) and outer parts of the envelope ($f_{\mathrm{out}}$;
$T$$<$90\,K).  For most species, the isotopic abundance ratios are
assumed to be fixed to the standard interstellar values to increase
the number of constraints used in the modelling.  The results from the
excitation analysis are presented in Fig.~\ref{chi2_jump} for several
species.  The reduced $\chi^2$ for the best fit models are generally
very good, $\sim$1, and significantly better than the constant
abundance models.  Typically, a jump of $\sim$100 in abundance is
derived, leading to abundances that are significantly higher than found in
quiescent molecular clouds and comparable to those found in the prototypical
hot core in Orion (Sect.~\ref{discussion}).

The jump models can only be applied to species for which a significant number
of lines are observed covering a wide range of excitation conditions.
These include H$_2$CO, CH$_3$OH, CH$_3$CN, H$_2$CS, SO and SO$_2$ 
(see the columns of $E_{\rm max}$ and $E_{\rm min}$ included in 
Table~\ref{abundances}).
For HC$_3$N and OCS, only lines from highly-excited levels have been
observed, so that for these molecules the values of $f_{\mathrm{out}}$
in the outer envelope are poorly constrained. For most simple
linear rotors such as HCN, HCO$^+$, CN, however, the observed lines arise
from levels below 90~K, so that no information on the inner warm part is
obtained. The only exception is SiO, where the combination of many
$^{28}$SiO and $^{29}$SiO lines allows a jump to be inferred.
Molecules such as HNCO, CH$_2$CO and H$_2$S for which only one or two lines are 
observed and where the emission mainly probes hot gas (Table~\ref{abundances})
only the inner part of the envelope were modelled. The significantly 
higher abundances obtained (Sect.~\ref{discussion}) compared with the 
constant abundance models illustrates the point that orders
of magnitude higher abundances can be derived if the emission is assumed to
originate only from the inner warm region.  

%
\begin{table}
   \caption[]{Various molecular abundances derived using a jump 
      in their fractional abundance, introduced at $T$$=$90\,K.}
      \label{jump_abundances} 
      $$
      \begin{array}{p{0.2\linewidth}cccccccccc}
	 \hline
     	 \noalign{\smallskip}
     	 \multicolumn{1}{c}{\mathrm{Molecule}} & 
	 \multicolumn{1}{c}{{f_{\mathrm{in}}(X)}^{\mathrm a}} & &
     	 \multicolumn{1}{c}{{f_{\mathrm{out}}(X)}^{\mathrm b}} & &
     	 \multicolumn{1}{c}{\chi^2_{\mathrm{red}}} & &
     	 \multicolumn{1}{c}{N} \\
     	 
	 \noalign{\smallskip}
	 \hline
	 \noalign{\smallskip}
	 HC$_3$N			      & \phantom{\mathrm{<}}1.0\times 10^{-9}\phantom{^0}  && \mathrm{<}1.0\times 10^{-10} 		      && \phantom{0}0.1 &&  \phantom{0}3    \\
     	 CH$_3$CN			      & \phantom{\mathrm{<}}7.5\times 10^{-9}\phantom{^0}  && \mathrm{<}8.0\times 10^{-11} 			 && \phantom{0}0.9 &&  \phantom{0}7  \\

	 C$_3$H$_2$			      & \mathrm{<}1.5\times 10^{-9}\phantom{^0}		&& \phantom{\mathrm{<}}1.6\times 10^{-11}	 && \phantom{0}5.7 &&  \phantom{0}6  \\
     	 CH$_3$C$_2$H			      & \phantom{\mathrm{<}}3.5\times 10^{-8}\phantom{^0}  && \mathrm{<}1.5\times 10^{-9}\phantom{^0} && \phantom{0}1.7 &&  \phantom{0}6  \\

     	 o-H$_2$CO			      & \phantom{\mathrm{<}}4.5\times 10^{-8}\phantom{^0}  && \phantom{\mathrm{<}}4.5\times 10^{-10}	 && \phantom{0}1.9 &&  \phantom{0}9  \\
     	 p-H$_2$CO			      & \phantom{\mathrm{<}}1.5\times 10^{-8}\phantom{^0}  && \phantom{\mathrm{<}}2.5\times 10^{-10}	 && \phantom{0}1.4 &&  \phantom{0}8  \\
     	 H$_2^{13}$CO			      & \phantom{\mathrm{<}}7.0\times 10^{-10}  && \phantom{\mathrm{<}}1.0\times 10^{-11}	 && \phantom{0}1.6 &&  \phantom{0}6  \\
     	 HDCO				      & \phantom{\mathrm{<}}2.0\times 10^{-8}\phantom{^0}  && \mathrm{<}9.0\times 10^{-11}	 && \phantom{0}0.9 &&  \phantom{0}4  \\
     	 CH$_3$OH			      & \phantom{\mathrm{<}}1.0\times 10^{-7}\phantom{^0}  && \phantom{\mathrm{<}}3.5\times 10^{-10}  && \phantom{0}1.2 &&      23       \\

     	 SO				      & \phantom{\mathrm{<}}2.5\times 10^{-7}\phantom{^0}  && \phantom{\mathrm{<}}3.5\times 10^{-9}\phantom{^0}	      && \phantom{0}1.8 &&  \phantom{0}9 \\
     	 SO$_2$ 			      & \phantom{\mathrm{<}}1.0\times 10^{-7}\phantom{^0}  && \phantom{\mathrm{<}}4.5\times 10^{-10}	 && \phantom{0}0.5 &&	    10 \\
     	 OCS				      & \phantom{\mathrm{<}}2.5\times 10^{-7}\phantom{^0}  && \mathrm{<}3.0\times 10^{-9}\phantom{^0} 	 && \phantom{0}1.5 &&  \phantom{0}7 \\
     	 o-H$_2$CS			      & \phantom{\mathrm{<}}3.0\times 10^{-9}\phantom{^0}  && \phantom{\mathrm{<}}1.0\times 10^{-10}	 && \phantom{0}1.4 &&  \phantom{0}6  \\
     	 p-H$_2$CS			      & \phantom{\mathrm{<}}2.5\times 10^{-9}\phantom{^0}  && \mathrm{<}4.0\times 10^{-11} 			 && \phantom{0}0.1 &&  \phantom{0}4  \\

     	 SiO				      & \phantom{\mathrm{<}}4.5\times 10^{-9}\phantom{^0}  && \phantom{\mathrm{<}}2.5\times 10^{-11}	 && \phantom{0}0.8 &&  \phantom{0}8 \\
     	 
     	 \noalign{\smallskip}
	 \hline
      \end{array}
      $$
      \smallskip

      \noindent
      $^{\mathrm a}$~Abundance in the inner, dense, and hot part of the envelope.\\
      \noindent
      $^{\mathrm b}$~Abundance in the cooler, less dense, outer part of the envelope.\\
\end{table}

The observations analyzed in this paper do not include the lowest
rotational transitions probing the coldest outer parts; thus,
so-called `anti-jump' models, in which the abundances are decreased
below a certain temperature due to freeze-out, cannot be tested,
except for the case of CO. There are also some molecules, e.g.,
C$_3$H$_2$, for which the jump-models give a worse $\chi^2$-fit than
the constant abundance models; such molecules are good candidates for
the `anti-jump' models if lower transitions are available.  In the
following, a few individual cases are described in more detail, before
discussing the general results.

\subsection{CO}
Carbon monoxide, CO, is difficult to destroy but relatively easy to
excite through collisions, even in the outer low-density and cold part
of the envelope.  Brightness maps of the $^{12}$CO molecular line
emission associated with \object{IRAS 16293--2422} reveal a complex
structure \citep{Walker88} indicating two bipolar outflows.
Interferometric BIMA observations suggest that the
$^{13}$CO($J$$=$1$\rightarrow$0) emission is also associated with the
outflow to some extent (Sch\"{o}ier et~al.\ 2002a, in prep.).
Single-dish observations of higher transitions of $^{13}$CO show no
direct evidence for tracing the outflow based upon the shape of their
line profiles (see Fig.~\ref{overlay}).  This is also the case for
lines from the less abundant C$^{17}$O and C$^{18}$O molecules.  Thus
these lines can possibly be used to trace the CO content in the
circumstellar envelope.

Assuming a constant abundance throughout the envelope the radiative
transfer calculations give abundances of 6.5$\times$10$^{-7}$,
6.2$\times$10$^{-8}$ and 1.6$\times$10$^{-8}$ for $^{13}$CO,
C$^{18}$O, and C$^{17}$O, respectively. The derived
C$^{18}$O/C$^{17}$O ratio is 3.9, in good agreement with typical
interstellar values \citep{Wilson94}. 
\citet{Jorgensen02} derived
$^{18}$O/$^{17}$O$=$3.6$\pm$1.1 for their survey of protostellar
objects. 
The $^{12}$CO abundance is
estimated to be 4.0$\times$10$^{-5}$ from C$^{17}$O assuming the
terrestrial ratio $^{12}$CO/C$^{17}$O=2500. This is in excellent
agreement with the value of 3.9$\times$10$^{-5}$ estimated from
$^{13}$CO using the interstellar value $^{12}$CO/$^{13}$CO=60.  In
addition, the upper limits obtained for the
$^{13}$C$^{17}$O($J$$=$2$\rightarrow$1) and
$^{13}$C$^{18}$O($J$$=$3$\rightarrow$2) line emission are also
consistent with these values.  In all, the consistency of the derived
values and quality of the model fits (Fig.~\ref{overlay} and
Table~\ref{abundances}) are reassuring.  Due to the high optical
depths in the observed $^{13}$CO lines, their profiles tend to be
somewhat broader than those from the less abundant isotopomers, and
their intensities are less sensitive to the assumed abundance.  The
observed lines are all in LTE, indicating that the derived abundances
are not sensitive to the adopted set of collisional rates.

\begin{figure} 
\centering
 \includegraphics[angle=-90,width=8cm]{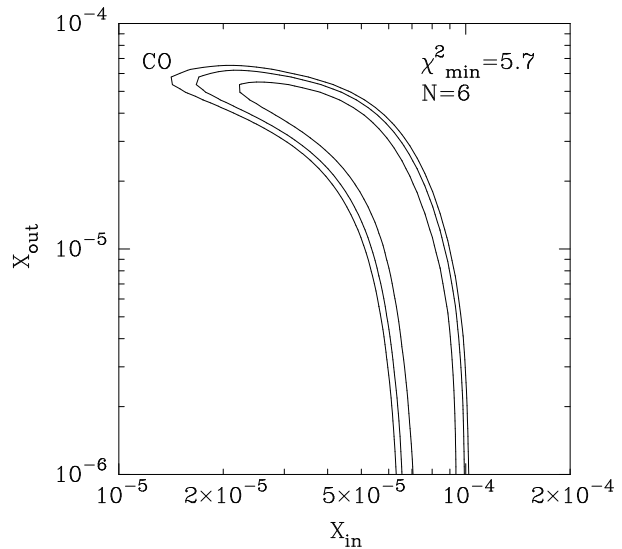}
 \caption{CO jump-model assuming an abrupt change in abundance at 20\,K using the 
 integrated intensities of the observed $^{13}$CO, C$^{18}$O, and C$^{17}$O line emission
 as constraints. Contours
 are drawn at $\chi^2_{\mathrm{min}}$$+$(2.3, 4.6, 6.2) indicating
 the 68\%, 90\%, and 95\% confidence levels, respectively. The
 number of observational constraints used, $N$, are also shown.  The
 quality of the best fit model can be estimated from the reduced
 chi-squared
 $\chi^{2}_{\mathrm{red}}$$=$$\chi^{2}_{\mathrm{min}}$/($N$$-$2).}  
 \label{frozen}
\end{figure}

The total inferred CO abundance is about a factor of two to four lower
than the value of $8\times 10^{-5}$ found in dark clouds
\citep{Frerking82} and $2\times 10^{-4}$ in warm regions
\citep{Lacy94}. A plausible explanation is that CO freezes out in the
cool external parts of the envelope. To simulate this situation, a
`jump' model with an abrupt decrease in the CO abundance at 20\,K
was introduced; this is a characteristic temperature below which
pure-CO ice can exist \citep{Sandford93}.  To compensate for this
freeze-out, the CO abundance in regions above 20\,K needs to be
raised.  In Fig.~\ref{frozen} the result of varying the CO abundance
in the inner and outer parts of the envelope is presented.  A maximum
allowed value for the CO abundance of gas above 20\,K is
$\sim$7$\times$10$^{-5}$ while at the same time the abundance in the
outer cooler envelope needs to be lowered to
$\lesssim$2$\times$10$^{-5}$, i.e., a depletion of about a factor
four.  However, in the present analysis a constant abundance model is
equally probable.  
The molecular line modelling performed in 
Sect.~\ref{dust_oh5} suggests that in the envelope around \object{IRAS 16293--2422},
$T_{\mathrm{gas}}$$\gtrsim$0.7$\times$$T_{\mathrm{dust}}$. This in
turn means that it is not possible to explain the apparently low CO
abundance with a significant decoupling of the gas temperature from
that of the dust.

For the similar modelling of the larger sample of class 0 and I objects,
\citet{Jorgensen02} found that the CO abundance in general was lower 
for the class 0 objects than the class I objects (average CO abundances of respectively
2.0$\times$10$^{-5}$ and 1.2$\times$10$^{-4}$). Further it was found that 
abundance jumps at 20\,K of more than a factor 3 could be ruled out in most 
cases and that constant fractional abundances over the temperature range 
covered by the CO rotational lines provided good fits. This lead to the suggestion 
that CO in the class 0 objects could be trapped in a porous ice matrix with 
H$_2$O from which it does not fully evaporate until at temperatures of $\sim$ 
60\,K. 
More observational constraints on isotopic CO
($J$$=$1$\rightarrow$0) as well as higher$-$$J$ CO lines are needed to
verify if CO is frozen out onto dust grains only at the lowest
temperatures or if a substantial fraction of CO evaporates more
gradually up to $\sim$90~K, as suggested by recent experiments of
CO--H$_2$O ice mixtures \citep{Collings02}.

In Fig.~\ref{overlay} the line profiles obtained from the constant
abundance model of the
CO emission are presented.  The asymmetry present in some of the
observed spectra is not possible to model using a static envelope and
requires the presence of a global velocity field, as discussed in
Sect.~\ref{infall}.

\subsection{H$_2$CO and CH$_3$OH}
H$_2$CO and CH$_3$OH are two particularly useful molecules to model in
detail, because of the relative complexity of the energy level
structure of asymmetric rotors and symmetric top molecules.  For
example, for H$_2$CO dipole transitions between various $K_p$-ladders
are not allowed while collisional transitions are, making these
transitions good probes of the temperature structure. In addition, the
wealth of available lines in the submillimetre regime makes them ideal
to probe various parts of the envelope
\citep{Mangum93,Dishoeck93,Ceccarelli00b}.  It is clear from
Table~\ref{jump_abundances} that the reduced $\chi^2$ values are
considerably lowered in the jump-model for an abundance contrast of
$\sim$60$-$100 for H$_2$CO and $\sim$50 for CH$_3$OH.  The H$_2$CO and
CH$_3$OH abundances in the cool outer part of the envelope are in line
with values in quiescent molecular clouds, whereas those in the inner
envelope are of the same order of magnitude as derived for some hot
cores (Sect.~\ref{discussion}).  Typical CH$_3$OH and H$_2$CO abundances in interstellar ices
are $\lesssim$2$\times$10$^{-6}$ and $\sim$10$^{-6}$
\citep{Dartois99, Keane01}, illustrating that the drastic increases
could be explained by ice evaporation from dust grains above 90~K.
For high-mass protostars, \citet{Tak00b} found a similar CH$_3$OH
abundance jump for some sources, but no evidence for significant jumps
in the H$_2$CO abundance. The H$_2$CO/CH$_3$OH ice abundance ratio is
sensitive to the atomic hydrogen density in the ice-forming region,
which could be different for low- and high-mass protostars.

When a significant number of transitions with varying excitation
conditions are observed the data have the potential to determine the
characteristic temperature at which the majority of the molecules are
evaporated. The sensitivity of the data presented here to the adopted
jump-temperature can thus be tested for H$_2$CO and CH$_3$OH.  
The H$_2$CO
models are not very sensitive to the jump temperature until it drops
below about 40\,K at which point the reduced $\chi^2$ increases
fast. It is found that a temperature of $\sim$50\,K gives the best fit
indicating that formaldehyde starts to evaporate at temperatures below
90\,K (see also Ceccarelli et al.\ 2001\nocite{Ceccarelli01}). In the case of methanol
(CH$_3$OH), the best fit is obtained for $\sim$90\,K whereas
temperatures below $\sim$50\,K give poor fits, indicating that this
molecule evaporates mainly at $\sim$90\,K.

The ortho-to-para ratio of H$_2$CO is not well constrained. Based on
the $\chi^2$-analysis, its lower limit is $\sim$0.9, with a best-fit
value of 2.5.  A further complication is that the ortho-to-para ratio
may vary, e.g.\ according to temperature, through the envelope. In the
present analysis it is not possible to confirm this.  Fixing the
ortho-to-para ratio to 2.5 throughout the envelope  increases the 
number of constraints used in the modelling and  enables
`jump-models' for H$_2$$^{13}$CO and HDCO to be inferred. From the
isotopic ratio, the derived $^{12}$C/$^{13}$C-ratio is $\sim$100 with
considerable uncertainty and fully consistent with the interstellar
value of 60 adopted elsewhere in this paper.  From deuterated
formaldehyde (HDCO) it is possible to estimate the D/H ratio and a
value $\sim$0.3 is derived.  This is significantly higher than what is obtained 
from DCO$^{+}$/HCO$^{+}$, DCN/HCN and DNC/HNC (see Table~\ref{abundances}),
and suggests a different scenario for H$_2$CO. The high degree of deuterium fractionation of H$_2$CO, further strengthened by the recent detection of D$_2$CO in this source \citep{Loinard00, Ceccarelli01}, is about five times larger than values obtained towards other low-mass protostars \citep{Roberts02} and 
dark clouds like TMC--1 and L134N  \citep{Turner01} but consistent with estimates for the Orion hot core \citep{Turner90}.
Due to the limited number of lines any radial variations in either the $^{12}$C/$^{13}$C-ratio or the
deuterium fractionation cannot be established.
\citet{Ceccarelli01} argue, based on spatially resolved emission, that H$_2$CO and its deuterated counterparts are formed mainly from grain-surface reactions in a previous cold, dark cloud phase. 
For HCO$^{+}$, HCN and HNC the degree of deuteration can be explained by gas-phase reactions at low temperatures \citep{Roberts02}.

\subsection{Vibrationally excited emission?}
For linear rotors such as CS and HCN, the observed lines cover only a
limited range in excitation energy. Much higher frequency data are
needed to probe the abundances of these molecules in the warm gas
through pure rotational lines in the vibrational ground state (e.g.,
Boonman et al.\ 2001\nocite{Boonman01}). Some lines of
vibrationally-excited molecules have been detected toward \object{IRAS
16293--2422}, however, in particular the CS($v$$=$1;$J$$=$7$\rightarrow$6) line.
Since this line originates from a level at about 1900\,K above ground,
it cannot be excited by collisions only. Instead, radiative excitation
by dust emission must play a role.

The addition of radiative excitation by dust is straightforward in the Monte Carlo
scheme if scattering is assumed to be negligible, which is the case at
the wavelengths of importance here: the CS fundamental vibrational
transition occurs around 8\,$\mu$m. In the absence of scattering events
only emission and absorption by the dust particles need to be
considered. The dust is assumed to locally emit thermal radiation
described by the dust temperature $T(r)$, according to Kirchhoff's
law. The model photons emitted by the dust are released together with
the other model photons and the additional opacity provided by the
dust is added to the line optical depth.

The CS abundance of 2.5$\times$10$^{-9}$ derived previously fails to
account for the observed CS($v$$=$1;$J$$=$7$\rightarrow$6) line
emission by many orders of magnitudes. Even a `jump-model' with the
abundance increased by a factor of 100 fails to account for all the
observed emission.

Recently, \citet{Highberger00} failed to detect any vibrationally
excited CS emission in lower $J$-transitions towards \object{IRAS
16293--2422} down to $T_{\mathrm{rms}}$$\sim$10\,mK. One possibility is
that the CS($v$$=$1;$J$$=$7$\rightarrow$6) line was mis-identified by
\citet{Blake94}. Their observations were obtained in dual sideband and
the CS line was supposed to reside in the lower sideband. However, in
the upper sideband two H$_2$CS lines
($J_{K_-,K_+}$$=$10$_{5,6}$$\rightarrow$9$_{5,5}$ and
$J_{K_-,K_+}$$=$10$_{5,5}$$\rightarrow$9$_{5,4}$) at 343.202331\,GHz
coincide with the position of the CS line in the lower sideband.  To
check the possibility that these transitions contribute significantly
to the observed intensity of the line, a LTE jump-model was run using
the parameters derived previously for H$_2$CS. It is indeed found that
these high lying energy levels ($E_{\mathrm u}$$=$418\,K) are
sufficiently excited to account for all of the flux observed in this
line.

\section{Discussion}
\label{discussion}

\subsection{Does IRAS 16293--2422 have a hot core?}

Hot cores have been defined for high-mass sources as small ($<$0.1\,pc), 
dense ($n_{\mathrm{H}_2}$$>$10$^7$\,cm$^{-3}$) and warm ($T$$>$100\,K) regions
\citep{Walmsley92,Kurtz00}.  The chemical signatures of hot cores are
high abundances of a wide variety of complex organic molecules and
fully hydrogenated molecules such as H$_2$O, NH$_3$ and H$_2$S. These
chemical characteristics are thought to arise from thermal evaporation
of ice mantles from the grains close to the protostar, followed by
rapid high-temperature gas-phase reactions for a period of 10$^4$$-$10$^5$
yr \citep{Charnley92}. In this scenario, the fully hydrogenated
species are `first generation' molecules produced by surface chemistry
on the grains, whereas complex organic molecules like HC$_3$N,
CH$_3$OCH$_3$ and HCOOCH$_3$ are `second generation' produced by
gas-phase chemistry between evaporated species.  To what extent do
these physical and chemical characteristics also apply to the 
low-mass object \object{IRAS 16293--2422}?

\begin{table*}
\caption[]{Summary of derived abundances in the envelope of \object{IRAS 16293--2422}.}
\label{summary} 
$$
\begin{array}{p{0.1\linewidth}cccccccccccccc}
\hline
\noalign{\smallskip}
 Molecule & &
 {f_{\mathrm{in}}(X)}^{\mathrm a}& &
 {f_{\mathrm{out}}(X)}^{\mathrm b}& &
 {\mathrm{L134N}}^{\mathrm c}& &
 {\mathrm{Orion}}^{\mathrm d}& &
 {\mathrm{L1157}}^{\mathrm e}& &
 {\mathrm{Ices}}^{\mathrm f}& &
 {\mathrm{Comets}}^{\mathrm g} \\

\noalign{\smallskip}
 & 
 &
 &
 &
 &
 &
 &
 &
 \multicolumn{1}{c}{\mathrm{hot\ core}} & &
 \multicolumn{1}{c}{\mathrm{outflow}} & &
 & &
 \multicolumn{1}{c}{\mathrm{Hale-Bopp}} \\
\noalign{\smallskip}
\hline
\noalign{\smallskip}
CO            &&   \cdots				 && \phantom{\mathrm{<}}4.0\times 10^{-5}\phantom{^0} && \phantom{\mathrm{<}}1\times 10^{-4}\phantom{^0} && \phantom{\mathrm{<}}1\times 10^{-4} \phantom{^0\mathrm{:}}&& 1\times 10^{-4}\phantom{^0} && \mathrm{(1-5)}\times 10^{-6}\phantom{^0} &&  1\times 10^{-5}\phantom{^0} \\
HCO$^+$       &&   \cdots				 && \phantom{\mathrm{<}}1.4\times 10^{-9}\phantom{^0} && \phantom{\mathrm{<}}8\times 10^{-9}\phantom{^0} && \phantom{\mathrm{<}}1\times 10^{-9} \phantom{^0\mathrm{:}}&& 3\times 10^{-8}\phantom{^0} && \cdots			&&	\cdots    \\ 

CN            &&   \cdots				 && \phantom{\mathrm{<}}8.0\times 10^{-11} && \phantom{\mathrm{<}}5\times 10^{-10}&& \cdots				   && 5\times 10^{-8}\phantom{^0} && \cdots			&&   \cdots \\
HCN           &&   \cdots				 && \phantom{\mathrm{<}}1.1\times 10^{-9}\phantom{^0} && \phantom{\mathrm{<}}7\times 10^{-9}\phantom{^0} && \phantom{\mathrm{<}}4\times 10^{-7} \phantom{^0\mathrm{:}}&& 5\times 10^{-7}\phantom{^0} && \mathrm{<}2\times 10^{-6}\phantom{^0}	&&  1\times 10^{-7}\phantom{^0} \\
HNC           &&   \cdots				 && \phantom{\mathrm{<}}6.9\times 10^{-11} && \phantom{\mathrm{<}}3\times 10^{-8}\phantom{^0} && \cdots				   && 5\times 10^{-8}\phantom{^0} && \cdots			&&  2\times 10^{-8}\phantom{^0} \\
HC$_3$N       && \phantom{\mathrm{<}}1.0\times 10^{-9}\phantom{^0}\phantom{^{\mathrm h}}  && \mathrm{<}1.0\times 10^{-10}	   && \phantom{\mathrm{<}}4\times 10^{-10} && \phantom{\mathrm{<}}2\times 10^{-9} \phantom{^0\mathrm{:}}&& 1\times 10^{-8}\phantom{^0} && \cdots			&&  1\times 10^{-8}\phantom{^0} \\
CH$_3$CN      && \phantom{\mathrm{<}}7.5\times 10^{-9}\phantom{^0}\phantom{^{\mathrm h}}  && \mathrm{<}8.0\times 10^{-11}	   && \mathrm{<}4\times 10^{-10}	   && \phantom{\mathrm{<}}2\times 10^{-8} \phantom{^0\mathrm{:}}&& \cdots	       && \cdots			&&  1\times 10^{-8}\phantom{^0} \\
HNCO          && {\phantom{\mathrm{<}}9.0\times 10^{-9}} ^{\mathrm h}\phantom{^0}  && \cdots				   && \cdots				   && \phantom{\mathrm{<}}6\times 10^{-9} \phantom{^0\mathrm{:}}&& \cdots	       && \mathrm{\sim}2\times 10^{-6}\phantom{^0} &&  5\times 10^{-8}\phantom{^0} \\

C$_2$H        &&  \cdots				 && \phantom{\mathrm{<}}2.2\times 10^{-10} && \phantom{\mathrm{<}}2\times 10^{-9}\phantom{^0} && \cdots				   && \cdots	       && \cdots			&&	  \cdots   \\
C$_3$H$_2$    &&  \cdots				 && \phantom{\mathrm{<}}1.6\times 10^{-11} && \phantom{\mathrm{<}}2\times 10^{-9}\phantom{^0} && \cdots				   && \cdots	       && \cdots			&&	  \cdots   \\
CH$_3$C$_2$H  && \phantom{\mathrm{<}}3.5\times 10^{-8}\phantom{^0}\phantom{^{\mathrm h}}  && \mathrm{<}1.5\times 10^{-9}\phantom{^0}	   && \mathrm{<}1\times 10^{-9}\phantom{^0}	   && \phantom{\mathrm{<}}1\times 10^{-9}\mathrm{:}\phantom{^0}				   && \cdots	       && \cdots			&&   \cdots   \\

H$_2$CO       && \phantom{\mathrm{<}}6.0\times 10^{-8}\phantom{^0}\phantom{^{\mathrm h}}  && \phantom{\mathrm{<}}7.0\times 10^{-10} && \phantom{\mathrm{<}}2\times 10^{-8}\phantom{^0} && \phantom{\mathrm{<}}1\times 10^{-8} \phantom{^0\mathrm{:}}&& 3\times 10^{-7}\phantom{^0} && \mathrm{(1-4)}\times 10^{-6}\phantom{^0} &&  5\times 10^{-7}\phantom{^0} \\ 
CH$_3$OH      && \phantom{\mathrm{<}}3.0\times 10^{-7}\phantom{^0}\phantom{^{\mathrm h}}  && \phantom{\mathrm{<}}3.5\times 10^{-10} && \phantom{\mathrm{<}}5\times 10^{-9}\phantom{^0} && \phantom{\mathrm{<}}2\times 10^{-7} \phantom{^0\mathrm{:}}&& 2\times 10^{-5}\phantom{^0} && \mathrm{(2-10)}\times 10^{-6}\phantom{^0}&&  1\times 10^{-6}\phantom{^0} \\ 
CH$_2$CO      && {\phantom{\mathrm{<}}5.0\times 10^{-8}}^{\mathrm h}\phantom{^0}  && \cdots				   && \mathrm{<}7\times 10^{-10}	   && \phantom{\mathrm{<}}3\times 10^{-10}\phantom{\mathrm{:}}				   && \cdots	       && \cdots			&&   \cdots   \\
HCOOH         && {\mathrm{<}8.0\times 10^{-9}}^{\mathrm h}\phantom{^0}		 && \cdots				   && \phantom{\mathrm{<}}3\times 10^{-10} && \phantom{\mathrm{<}}8\times 10^{-10} \phantom{\mathrm{:}}&& \cdots	       && \mathrm{(2-10)}\times 10^{-7}\phantom{^0}&&  5\times 10^{-8}\phantom{^0} \\
CH$_3$CHO     && {\mathrm{<}2.0\times 10^{-9}}^{\mathrm h}\phantom{^0}		 && \cdots				   && \phantom{\mathrm{<}}6\times 10^{-10} && {\lesssim}2\times 10^{-10}\phantom{\mathrm{:}}				   && \cdots	       &&     \cdots			&&	  \cdots   \\
CH$_3$OCH$_3$ && {\mathrm{<}4.0\times 10^{-8}}^{\mathrm h}\phantom{^0}		 && \cdots				   && \cdots				   && \phantom{\mathrm{<}}1\times 10^{-8}\phantom{^0\mathrm{:}}				   && \cdots	       &&     \cdots			&&   \cdots	  \\ 
HCOOCH$_3$    && {\mathrm{<}6.0\times 10^{-8}}^{\mathrm h}\phantom{^0}		 && \cdots				   && \cdots				   && \phantom{\mathrm{<}}1\times 10^{-8} \phantom{^0\mathrm{:}}&& \cdots	       &&   \cdots			&& 4\times 10^{-6}\phantom{^0}  \\

CS            &&  \cdots				 && \phantom{\mathrm{<}}3.0\times 10^{-9}\phantom{^0} && \phantom{\mathrm{<}}1\times 10^{-9}\phantom{^0} && \phantom{\mathrm{<}}1\times 10^{-8}\phantom{^0\mathrm{:}} && 2\times 10^{-7}\phantom{^0} &&   \cdots			&&  \cdots \\

SO            && \phantom{\mathrm{<}}2.5\times 10^{-7}\phantom{^0}\phantom{^{\mathrm h}}  && \phantom{\mathrm{<}}3.5\times 10^{-9}\phantom{^0} && \phantom{\mathrm{<}}6\times 10^{-9}\phantom{^0} && \phantom{\mathrm{<}}5\times 10^{-8}\phantom{^0\mathrm{:}} && 3\times 10^{-7}\phantom{^0} &&   \cdots			&& 1\times 10^{-7}\phantom{^0}  \\
SO$_2$        && \phantom{\mathrm{<}}1.0\times 10^{-7} \phantom{^0}\phantom{^{\mathrm h}} && \phantom{\mathrm{<}}4.5\times 10^{-10} && \phantom{\mathrm{<}}3\times 10^{-9}\phantom{^0} && \phantom{\mathrm{<}}6\times 10^{-8} \phantom{^0\mathrm{:}}&& 5\times 10^{-7}\phantom{^0} &&   \cdots			&& 1\times 10^{-7}\phantom{^0}  \\
OCS           && \phantom{\mathrm{<}}2.5\times 10^{-7}\phantom{^0}\phantom{^{\mathrm h}}  && \mathrm{<}3.0\times 10^{-9}\phantom{^0}	   && \phantom{\mathrm{<}}7\times 10^{-9}\phantom{^0} && \phantom{\mathrm{<}}5\times 10^{-8} \phantom{^0\mathrm{:}}&& 2\times 10^{-7}\phantom{^0} && 2\times 10^{-7}\phantom{^0}		&& 2\times 10^{-7}\phantom{^0}  \\
HCS$^+$       &&  \cdots				 && \phantom{\mathrm{<}}2.4\times 10^{-11} && \phantom{\mathrm{<}}1\times 10^{-10} && \cdots				   && 3\times 10^{-9}\phantom{^0} &&    \cdots			&&  \cdots	\\
H$_2$CS       && \phantom{\mathrm{<}}5.5\times 10^{-9}\phantom{^0}\phantom{^{\mathrm h}}  && \phantom{\mathrm{<}}1.0\times 10^{-10} && \cdots				   && \cdots				   && \cdots	       &&    \cdots			&& 1\times 10^{-8}\phantom{^0}  \\
H$_2$S        && \phantom{\mathrm{<}}9.0\times 10^{-8}\phantom{^0}\phantom{^{\mathrm h}}  && \cdots				   && \phantom{\mathrm{<}}8\times 10^{-10} && \mathrm{<}1\times 10^{-7}\phantom{^0\mathrm{:}}	   && 4\times 10^{-7}\phantom{^0} &&    \cdots			&& 7\times 10^{-7}\phantom{^0}  \\ 

SiO           && \phantom{\mathrm{<}}4.5\times 10^{-9}\phantom{^0}\phantom{^{\mathrm h}}  && \phantom{\mathrm{<}}2.5\times 10^{-11} && \mathrm{<}4\times 10^{-12}	   && \phantom{\mathrm{<}}6\times 10^{-8} \phantom{^0\mathrm{:}}&& 7\times 10^{-8}\phantom{^0} &&    \cdots			&&  \cdots	\\

   \noalign{\smallskip}
   \hline
\end{array}
$$
\smallskip

\noindent
$^{\mathrm a}$ Abundance in warm and dense inner part of the envelope $\leq$150\,AU in radius around IRAS 16293--2422\\
\noindent
$^{\mathrm b}$ Abundance in cooler, less dense outer part of the envelope around IRAS 16293--2422\\
\noindent
$^{\mathrm c}$ From \citet{Ohishi92}, updated using values from \citet{Dickens00} for L134N pos. C \\
\noindent
$^{\mathrm d}$ From \citet{Dishoeck98} (and references cited); otherwise  from \citet{Sutton95} \\
\noindent
$^{\mathrm e}$ From \citet{Bachiller97} at position B2 assuming CO/H$_2$$=$10$^{-4}$ \\
\noindent
$^{\mathrm f}$ {Based on \citet{Ehrenfreund00} assuming H$_2$O(ice)/H$_2$$=$5$\times$10$^{-5}$. 
  	       The ranges reflect the values found for different sources. The value for HNCO is derived from OCN$^-$,
  	       assuming this ice evaporates as HNCO}\\
\noindent
$^{\mathrm g}$ From \citet{Bockelee00} for comet Hale-Bopp at 1\,AU, assuming H$_2$O/H$_2$$=$5$\times$10$^{-5}$\\
\noindent
$^{\mathrm h}$ Assuming that the observed emission or upper limits apply to the inner warm region only\\
\end{table*}

The dust radiative transfer modelling performed in Sect.~\ref{dustres}
clearly indicates that there is evidence for the presence of hot and
dense material within $\sim$150\,AU of the protostar. Moreover, this
material is most probably in a state of collapse towards the
protostar. The need for introducing drastic jumps in the abundances of
some `first generation' species like H$_2$CO and CH$_3$OH at the
evaporation temperature of the ices ($\sim$90\,K) further suggests
that the ice mantles are liberated in the hot inner
regions. Table~\ref{summary} compares the abundances $f_{\rm in}$ of
these molecules with those found in interstellar ices and in high-mass
hot cores.  The table also includes the abundances found in comet
Hale-Bopp, which may be representative of interstellar ices. Although
the \object{IRAS 16293--2422} abundances are still up to an order of
magnitude lower than those found in typical ices, they are comparable
to those observed in high-mass hot cores. The high degree of deuterium
fractionation measured for these molecules
\citep{Dishoeck95,Ceccarelli01} is consistent with this
interpretation, and probably results from a combination of gas-phase
deuterium fractionation in the cold pre-collapsing cloud and
grain-surface reactions \citep{Tielens83, Roberts00}.

The high abundances of the sulfur-bearing species of $\sim$10$^{-7}$
may also fit with this scenario.  Molecules like SO, SO$_2$, OCS and
H$_2$CS are all predicted to be drastically enhanced by the injection
of H$_2$S into the hot core region \citep{Charnley97}, but may also
be present in the grain mantles themselves. H$_2$S
is detected toward \object{IRAS 16293--2422}, but because only a single
line is observed, the jump model cannot be uniquely constrained. If
all H$_2$S emission is assumed to come from the hot inner region, its
inferred abundance is $\sim$1$\times$10$^{-7}$.
This relatively high H$_2$S abundance is consistent with 
values found in a survey  of massive hot cores \citep{Hatchell98b} 
but an order of magnitude lower than the estimates for Orion \citep{Minh90}.

The physical characteristics in \object{IRAS 16293--2422} are thus
consistent with those of a typical `hot core' as observed towards
many high mass protostars, except for a large change in physical scale.  
The remaining question is whether the evaporated ices
have driven a similarly complex organic chemistry in this low-mass
protostar. Among potential `second generation' products, HC$_3$N and
CH$_3$CN are observed in fairly large amounts with some evidence for
jumps in their abundances. However, the \object{IRAS 16293--2422}
spectra do not show the wealth of spectral features due to other
complex organics -- although the confusion limit is clearly not yet
reached. The limits on the abundances of molecules such as
CH$_3$OCH$_3$ and HCOOCH$_3$, derived if it is assumed 
that these molecules are located only in the warm inner part of the envelope,
are higher than those
found for high-mass protostars (Table~\ref{summary}). Deep integrations 
down to $\sim$5-10\,mK are needed to verify the presence of these complex 
organic species.

Typically, species such as SO$_2$, OCS, HC$_3$N and CH$_3$CN are
assumed to be `second generation' products, in which case their
abundance ratios can be used as `chemical clocks' to determine the
time since the evaporation of the ice mantles. For the density and
temperature prevailing in the inner parts of the envelope around
\object{IRAS 16293--2422} an age of $\sim$10$^4$\,yr is inferred from
such chemical models, a value that is rather uncertain and sensitive
to the cosmic ray ionization rate (e.g., Charnley
1997\nocite{Charnley97}; Hatchell et al.\ 1998\nocite{Hatchell98b}).
However, if the velocity field derived from the dust modelling
for the Shu infall model is correct, the transit time for grains and
molecules through the warm, dense region surrounding \object{IRAS
16293--2422} is only several hundred years.  This is not sufficient
time for extensive second generation processing to result, and would
naturally explain the lack of large, complex organic species toward
this source if their abundances are indeed found to be very low by
subsequent deep searches.  In such a scenario, the observed
abundances would provide a unique opportunity to trace the chemical
richness derived from previous stages of grain mantle and gas phase
chemistry at sensitivities much higher than those achieved via
infrared spectroscopy of icy grains.

\begin{figure} 
\centering
 \includegraphics[angle=-90,width=8cm]{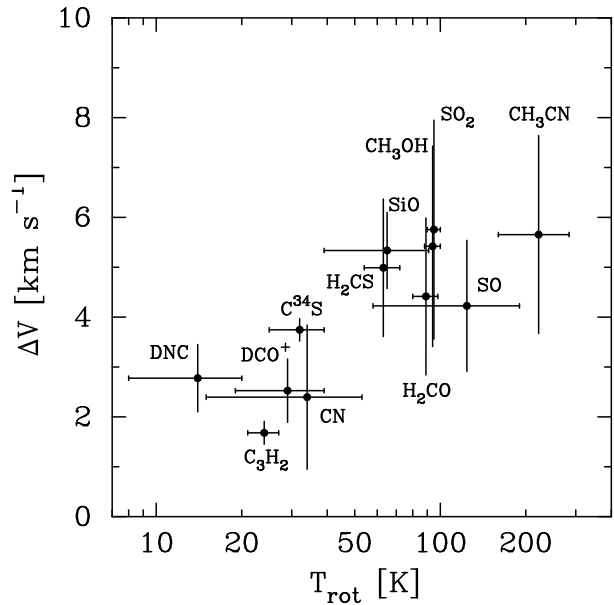}
 \caption{Observed line widths ($\Delta V$) of different molecules 
   as functions of their estimated rotational excitation temperature 
   ($T_{\mathrm{rot}}$).}  
 \label{rottemp}
\end{figure}

\subsection{Alternative scenarios}
The above discussion demonstrates that the \object{IRAS 16293--2422}
data can be consistently interpreted with a `hot core' in which the
ices are liberated from the grains by thermal evaporation above
90\,K. However, one problem is that the size of the `hot core' is only
$\sim$150\,AU in radius, comparable to the size(s) of the circumstellar
disk(s), and much smaller than the binary separation.  On this scale,
departures from spherical symmetry can be expected to play a role and
the disk can shadow part of the inner envelope, keeping it much
colder. Moreover, the \object{IRAS 16293--2422} envelope is known to be
traversed by outflows. To what extent could these processes affect the
interpretation? The data clearly indicate jumps in the abundances of
`first generation' molecules known to be present in interstellar ices,
but could these icy mantles also be removed by alternative mechanisms?

\citet{Blake94} and \citet{Dishoeck95} proposed that
grain-grain collisions in the turbulent shear zones where the outflow
interacts with the envelope can also be effective.  This mechanism is
observed for the class 0 protostar L1157, where the outflow
interaction can be spatially separated from the immediate protostellar
environment with single-dish telescopes. The inferred abundances by
\citet{Bachiller97} are included in Table~\ref{summary} and are seen to 
also be close to the values found in the inner region of \object{IRAS
16293--2422}, especially for the sulfur-bearing species.

An interpretation in which the ices are liberated by mild shocks or
turbulence rather than thermal evaporation has two observational
consequences.  First, the lines of the `first generation' ice mantle
species are expected to be wider and have a different velocity
structure than those of molecules located predominantly in the
quiescent outer envelope.  Figure~\ref{rottemp} shows the observed
line widths of different molecules as functions of their excitation
temperature, where the values are taken from \citet{Blake94} and
\citet{Dishoeck95}.  Molecules with clear `jumps' in their abundances
(SO$_2$, CH$_3$OH, CH$_3$CN) have larger line widths and higher
excitation temperatures than molecules which trace the outer envelope
(CN, C$_2$H, DNC, DCO$^+$).  
 The line widths observed towards \object{IRAS 16293--2422} are
significantly wider than what is typically observed for class 0
sources ($\lesssim$1\,km\,s$^{-1}$ for C$^{18}$O and C$^{17}$O; 
J{\o}rgensen et~al.\  2002\nocite{Jorgensen02})
The observed widths up to 8\,km\,s$^{-1}$
could, however, possibly be explained by the infall model with thermal
evaporation (Fig.~\ref{tdust}) and are not clear-cut evidence for
association with the molecular outflow, illustrating the difficulty in
disentangling the contributions from various velocity components.
Second, the spatial distribution of the `first generation' molecules
will be different. Molecules produced by thermal evaporation should be
located within a $\sim$150\,AU (1$\arcsec$) radius, whereas in the
shock scenario they are expected to coat the walls of the
outflow(s). Such an `X-type' geometry can extend over a much larger
region, even though the total mass of warm gas may be comparable to
that in the first model.  Chemically, there is expected to be little
difference between the two scenarios, except perhaps in the `second
generation' products if the liberation by shocks occurs in lower
density or temperature gas.  Observations at sub-arcsec resolution
with the Smithsonian SubMillimeter Array (SMA) and the Atacama Large
Millimeter Array (ALMA) are needed to distinguish these scenarios.

 Recently, \citet{Viti01} have suggested, based on chemical modelling,
that some molecular abundance ratios are strongly affected by the presence 
of a shock. In these models, the shocks not only liberate the ice mantles but also 
drive high-temperature ($\sim$2000\,K) reactions.
In particular the combination of HCO/H$_2$CO and NS/CS ratios 
may be suited for tracing the dynamical history of a hot core.
For the HCO/H$_2$CO ratio we derive an upper limit of $\sim$1, about an order of magnitude
larger than any model predictions, whereas no NS data are available. 
Thus, the abundances derived from the 
data set cannot be used to constrain their origin in the inner hot region of 
\object{IRAS 16293--2422}.

\section{Conclusions}
The continuum emission radiated by dust grains in the circumstellar
environment of recently born stars can be used, when supplemented by a
detailed radiative transfer analysis, to derive the physical
properties of such dusty envelopes. Observational constraints in the
form of both the SED and resolved radial brightness distributions are
required for a successful model. For the protostellar object
\object{IRAS 16293--2422} it is possible to model the observational data
using a single power-law distribution for the density
structure. However, the molecular line profiles suggest that parts of
the envelope are in a state of collapse. A dynamical age of
$\sim$1-3$\times$10$^4$\,yr is derived with an annual mass accretion
rate of roughly 4$\times$10$^{-5}$ M$_{\odot}$ yr$^{-1}$.

Once the physical structure of \object{IRAS 16293--2422} is known, it
is possible to determine its chemical properties through detailed
radiative transfer modelling of the observed molecular line emission.
 The abundances and quality of the fits for molecules like CO and CS
further strengthen the adopted physical model.
While the emission from some molecules is well reproduced assuming a
constant fractional abundance throughout the envelope, other species
require a steep abundance gradient to be introduced at typically
90~K. The presence of such a jump for molecules like H$_2$CO and
CH$_3$OH is interpreted as evidence of thermal evaporation of ices in
the inner dense and hot regions of the envelope. Other molecules like
HC$_3$N, CH$_3$CN and several sulfur-bearing molecules also require
such drastic jumps in their abundance distributions.  In high mass
protostars, these molecules are typically cited as evidence of a 
rapid gas-phase chemistry initiated by the evaporation of the ices.
The `hot core' region in \object{IRAS 16293--2422} is, however, very 
small, only $\sim$150\,AU in radius and comparable to the size of 
the circumstellar disk(s).  This large change in physical size
and the organized velocity field surrounding low mass protostars lead
to chemical time scales that only a small fraction of the dynamical age.
Thus, it may be difficult for a full `hot core' chemistry leading to
complex species such as CH$_3$OCH$_3$ to develop in class 0 objects,
even though the appropriate physical conditions are present.
Alternative scenarios in which ices are liberated by grain-grain
collisions in turbulent shear zones associated with the outflows need
to be tested by higher angular resolution observations.

Whatever their precise origin, the molecules located in the inner
envelope of \object{IRAS 16293--2422} can be incorporated into the
growing circumstellar disk(s) and become part of the material from
which planetary bodies are formed. The molecular abundances derived
here should provide an accurate reference point for comparison with
the growing amount of data on protoplanetary disks and icy solar
system objects such as comets.

\begin{acknowledgements}
The authors are grateful to R. Stark and S.\ Doty for useful discussions.
 The referee J. Hatchell is thanked for comments that helped to
improve the paper.
This research was supported by the Netherlands Organization for
Scientific Research (NWO) grant 614.041.004, the Netherlands Research School
for Astronomy (NOVA) and a NWO Spinoza grant.
This article made use of data obtained through the JCMT archive as Guest User 
at the Canadian Astronomy Data Center, which is operated by the Dominion
Astrophysical Observatory for the National Research Council of Canada's
Herzberg Institute of Astrophysics. 

\end{acknowledgements}

\bibliographystyle{aa}

\end{document}